\documentclass[aps,prd,nofootinbib]{revtex4}
\usepackage{graphicx}
\usepackage{hyperref}
\usepackage{slashed}
\usepackage{color}
\usepackage{subfigure}
\usepackage{ulem}
\usepackage{bm}
\usepackage{multirow}
\usepackage{hyperref}

\begin{document}

\title {Production of double heavy quarkonia at super $Z$ factory}

\author{Qi-Li Liao$^{a}$\footnote{xiaosueer@163.com}}
\author{Jun Jiang$^{b}$\footnote{jiangjun87@sdu.edu.cn, corresponding author}}
\author{Yu-Han Zhao$^{b}$\footnote{zyh329@mail.sdu.edu.cn}}

\affiliation{$^a$Chongqing College of Mobile Telecommunications, Chongqing 401520, China\\
$^b$School of Physics, Shandong University, Jinan 250100, Shandong, China}

\date{\today}

\begin{abstract}
Within the color singlet model, we calculate the exclusive production of double charmonia, double bottomonia, and double $B_c$ mesons at future super $Z$ factory.
The two heavy quarkonia or $B_c$'s are either two S-wave Fock states ($^1S_0, ~^3S_1$), or one S-wave and one P-wave states ($^1P_1,~^3P_J~(J=0,1,2)$).
The top three $Z^0$ propagated channels in cross sections for double charmonia are
$\sigma{(J/\psi+h_c)}_{Z^0}=(454.0^{+39.1}_{-59.2})\times10^{-5} fb,
\sigma{(\eta_c+\chi_{c2})}_{Z^0}=(284.2^{+22.2}_{-35.3})\times10^{-5} fb$, and
$\sigma{(\eta_c+\chi_{c0})}_{Z^0}=(137.0^{+9.9}_{-16.4})\times10^{-5} fb$.
For double bottomonia, they are
$\sigma{(\eta_b+\Upsilon)}_{Z^0}=(428.6^{+67.9}_{-60.9})\times10^{-4} fb,
\sigma{(\Upsilon+\chi_{b2})}_{Z^0}=(230.6^{+23.6}_{-22.2})\times10^{-4} fb$, and
$\sigma{(\Upsilon+\Upsilon)}_{Z^0}=(119.9^{+18.4}_{-16.6})\times10^{-4} fb$.
For double $B_c$ mesons, they are
$\sigma{(B_c^{*+}+B_c^{*-})}_{Z^0}=(1150^{+87}_{-84})\times10^{-3} fb$,
$\sigma{(B_c^{*+}+\chi_{bc2}^-)}_{Z^0}=(1133^{-16}_{+22})\times10^{-3} fb$, and
$\sigma{(\eta_{bc}^++B_c^{*-})}_{Z^0}=(634.6^{+47.8}_{-45.8})\times10^{-3} fb$.
Here the uncertainties come from the varying masses of constituent heavy quarks, which bring up to 20\% corrections.
The cross sections of double $B_c$ mesons are roughly one order of magnitude larger than those of the double bottomonia, and two orders of magnitude larger than those of the double charmonia.
To make it helpful for experimental study, we present the total cross sections $\sigma$ as functions of CM energy $\sqrt{s}$, $\sigma$ as functions of the renormalization scale $\mu$, the angle distributions $d\sigma/dcos\theta$, and the $p_T$ distributions $d\sigma/dp_{t}$.
We also find that the initial state radiation can bring about 30\%$\sim$40\% suppresions when 1\%$m_Z$ energy is losing, and
cross sections can increase by about $2\sim3$ times or decrease by an order of magnitude when adopting different potential models which becomes the major source of uncertainty.
The numerical results show that it might be not optimistic for the experimental observation, but it is still far from excluded at the FCC-ee and also the CEPC running in the $Z$ factory mode.
\end{abstract}

\maketitle

%%%%%%%%%%%%%%%%
%%%%%%%%%%%%%%%%
\section{Introduction}
%%%%%%%%%%%%%%%%
%%%%%%%%%%%%%%%%
%$Z$-FACTORY\\
The Large Hadron Collider (LHC) provides a direct probe of the high energy frontier of particle physics.
We also have the BESIII experiment at the BEPC running at the $\tau$-charm energy region and the Belle II experiment at the super KEKB running as a $B$ factory.
As a supplementary, we further need another electron-positron collider to make precise study on the $W^{\pm},~Z^0$, and Higgs physics running at proper energy regions.
The Circular Electron Positron Collider (CEPC) provides three operation modes, corresponding to the Higgs factory running at center-of-momentum (CM) energy $\sqrt{s}=240$ GeV, the $Z$ factory at $\sqrt{s}=91.2$ GeV, and the $WW$ threshold scan around $\sqrt{s}\sim160$ GeV.
For the $Z$-factory mode, the CEPC would have the $Z^0$ event yields of $7\times10^{11}$ \cite{CEPCStudyGroup:2018ghi}.
The first stage of Future Circular Collider (FCC) is an electron-positron collider (FCC-ee), which would operate at the $Z$ resonance for four years.
It has even larger $Z^0$ event yields of $7\times10^{12}$, since its total integrated luminosity is about nine times as that of CEPC \cite{Agapov:2022bhm}.
The future electron-positron colliders would provide nice platforms for the precise study on $Z^0$ physics.
Note that, the Electron Positron Linear Collider also has the similar GigaZ running mode \cite{ECFADESYLCPhysicsWorkingGroup:2001igx}, and a super $Z$ factory has also been proposed by another Chinese group \cite{jz}.

%HEAVY QUARKONIUM\\
The production of heavy quarkonium is a multiscale problem for probing the quantum chromodynamics (QCD) theory.
The heavy quarkonium is a bound state consisting of a heavy quark and a heavy antiquark, i.e. the $Q\bar{Q}^\prime$ ($Q,~Q^\prime = c$ or $b$ quarks) bounding systems.
The production of the heavy quarkonium has three different momentum scales: the heavy quark mass $\sim m_Q$, the momentum of the heavy quark or antiquark in the quarkonium rest frame $\sim m_Q v$, and the kinetic energy of the heavy quark or antiquark $\sim m_Q v^2$.
Here, $v$ is the relative velocity between the constituent heavy quark and antiquark in the quarkonium rest frame.
The typical values are $v^2=0.1$ for $b\bar{b}$ quarkonium, and $v^2=0.3$ for $c\bar{c}$ quarkonium.
Because of $m_Q \gg m_Q v \gg m_Q v^2$, the creation of the constituent $Q\bar{Q}^\prime$ pair can be calculated perturbatively, while the hadronization of $Q\bar{Q}^\prime$ pair to a physical color-singlet quarkonium is nonperturbative.
So the production of heavy quarkonium has been a hot topic on exploring the QCD at both the perturbative and nonperturbative energy regions, which provides rich physics on strong interation.
For more information on the current status of heavy quarkonium, please refer to some reviews \cite{Brambilla:2010cs,Andronic:2015wma,Chung:2018lyq,Chen:2021tmf,Chapon:2020heu}.

%NRQCD & CSM\\
Considering the fact that the constituent heavy quark and antiquark inside the quarkonium have the nonrelativistic nature, we can use the nonrelativistic QCD (NRQCD) factorization framework \cite{nrqcd1,nrqcd2} to factor the production of heavy quarkonium into the perturbative creation of the constituent $Q\bar{Q}^\prime$ pair and the nonperturbative hadronization of the heavy quark pair.
The NRQCD factorization formulation consists of the short-distance coefficients and long-distance matrix elements.
The short-distance coefficients stand for the hard creation of the heavy constituent \textbf{$Q\bar{Q}$} pair and can be calculated perturbatively with the method of the Feynman diagrams.
The long-distance matrix elements stand for the nonperturbative hadronization of the binding \textbf{$Q\bar{Q}$} pair with definite $J^{PC}$ quantum numbers into the physical heavy quarkonium.
Our calculation are in the color singlet model (CSM) \cite{Ellis:1976fj,Carlson:1976cd,Chang:1979nn}, which require the intermediate \textbf{$Q\bar{Q}$} pair and the final quarkonium have the same quantum numbers.
Under this assumption, the long-distance matrix elements then can be related to the radial wave functions (or their derivatives) of the heavy quarkonium at the origin, which can be extracted from the decays of the quarkonim and can be calculated within the potential models \cite{wgs,pot2,Richardson:1978bt,IO,Ikhdair:2003ry,Chen:1992fq,Eichten:1978tg,Eichten:1979ms,Eichten:1980mw,Eichten:1995ch} or potential NRQCD \cite{Brambilla:1999xf} or lattice QCD \cite{Bodwin:1996tg}.
Thus there are no free nonperturbative parameters in CSM.

%TRACE TECHNOLOGY\\
In the calculation of the short-distance coefficients in the NRQCD framework, the squared amplitudes become complicated and lengthy especially for the production of the $P$-wave quarkonium.
To simplify the calculation, we adopt the ``improved trace technology" \cite{cjx,lxz,Yang:2011ps,wbc1,Liao:2012rh,Liao:2021ifc}, which is based on the helicity amplitude method.
In which, the amplitudes have been expressed directly as the linear combinations of independent Lorentz structures before the polarization sum.
This method helps a lot to overcome the problem of time-consuming calculation for the production of the $P$-wave quarkonium.

%RELATED WOKRS\\
The production of double heavy quarkonia has been studied extensively both at the LHC and $B$ factories.
The hadronic production of double $J/\psi$ has always been a hot topic because $J/\psi$ can be very easy to be detected by its leptonic decays and this channel can be used to explore the distribution of gluons in a proton at the LHC \cite{Schafer:2019ynn,Scarpa:2019fol,He:2019qqr,Qiao:2002rh,Qiao:2009kg,Lansberg:2013qka,Lansberg:2020rft,Lu:2021gxf}.
In addition, the photoproduction of double $J/\psi$ \cite{Xue-An:2018wat}, the production of double heavy quarkonia through diffractive interactions \cite{BrennerMariotto:2018eef}, the production of double heavy quarkonia through photon-photon interaction \cite{Chen:2020dtu,Yang:2020xkl}, and the hadronic production of double $B_c$ mesons \cite{Li:2009ug} have also been studied.
The production of double heavy quarkonia through electron-positron annihilation has been widely explored at $B$ factories.
In particular, the production of $J/\psi + \eta_c$ at $B$ factories once challenged the NRQCD framework.
At the leading order (LO) within NRQCD formulation, its total cross section is about $2\sim 6 ~ fb$ \cite{Braaten:2002fi,Liu:2002wq,Hagiwara:2003cw}.
However, the measurements at $B$ factories by the Belle and BaBar collaborations show that the cross section is much bigger, around $20 ~fb$ \cite{Belle:2002tfa,BaBar:2005nic}.
This large discrepancy between theory and experiment can be reduced by the QCD next-to-leading order (NLO) corrections \cite{Zhang:2005cha,Gong:2007db} and the relativistic corrections in the NRQCD \cite{Braaten:2002fi,He:2007te,Bodwin:2007ga}.
Recently, this issue has been solved by the QCD next-to-next-to-leading corrections \cite{Feng:2019zmt}, which gives consistent estimate with the BaBar measurement.

%THIS MANUSCRIPT\\
The future $Z$ factory is another platform to study the production of double heavy quarkonia through the $Z^0$ boson decays.
Although these exclusive processes are rare, we have the advantages of the vast $Z^0$ events and the clear background provided by the fully reconstructed quarkonia at the electron positron collider running as a $Z$ factory.
This provide us another opportunity of studying the interplay between the perturbative and nonperturbative regimes in the hadronization of heavy quark pair into quarkonia.
The production of the double charmonia at LO at such $Z$ factory is discussed in Refs. \cite{gxz1,Likhoded:2017jmx}.
The QCD NLO corrections to the production of double charmonia are also studied \cite{Berezhnoy:2021tqb,Luo:2022ugd}, which shows that NLO corrections are also important near the $Z^0$ mass pole.
The QCD NLO corrections to the paired $S$-wave $B_c^{(*)}$ production are obtained, where the NLO corrections are small at $Z^0$ pole \cite{Berezhnoy:2016etd}.
The associated $S$-wave charmonium-bottomonium production at LO at $Z^0$ pole are explored in Ref. \cite{Belov:2021ftc}.
In the present paper, we shall study the production of double charmonia, double bottomonia, and double $B_c$ mesons at the future $Z$ factory, where the case of two $S$-wave ($^1S_0, ~^3S_1$) states and the case of one $S$-wave and one $P$-wave ($^1P_1, ~^3P_J$ ($J=0,1,2$)) states are considered.
To make our work more helpful in future experimental researches, we shall discuss the dependences of cross sections on the center-of-momentum energy and the running renormalization scale, the differential angle distribution, the transverse momentum distribution, and the uncertainties caused by the masses of constituent heavy quarks and the radial wave functions (or their derivatives) of the heavy quarkonium at the origin.

%OUTLINES\\
The rest of this paper is organized as follows.
In Sec. II, we introduce the prescription of the production of double heavy quarkonium within the CSM framework.
In Sec. III,  we first present the total cross sections, then discuss their differential distributions and the uncertainties.
Sec. IV is reserved for a summary.

%%%%%%%%%%%%%%%%
%%%%%%%%%%%%%%%%
\section{Formulations}
%%%%%%%%%%%%%%%%
%%%%%%%%%%%%%%%%
%
\begin{figure}
\includegraphics[width=0.4\textwidth]{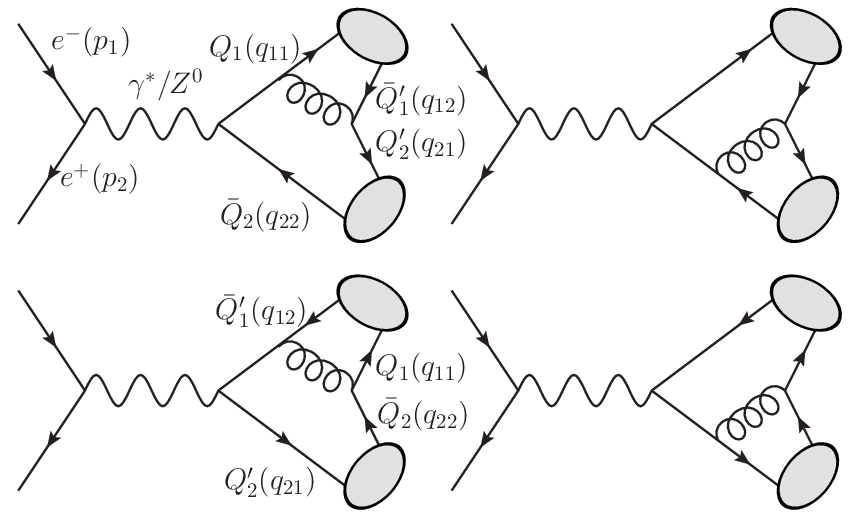}
\caption{Feynman diagrams for electron positron annihilation into double heavy quarkonia or $B_c$'s through the virtual photon and $Z^0$ boson.}
\label{feyn1}
\end{figure}
The Feynman diagrams for the exclusive production of double heavy quarkonia or $B_c$ mesons in $e^-(p_1) e^+(p_2) \to \gamma^*/Z^0 \to H_1[n](q_1) + H_2[n^\prime](q_2)$ are displayed in Fig.~\ref{feyn1}, where $H_{1(2)}$ represent the heavy quarkonia or $B_c$'s and $[n^{(\prime)}]$ represent the spin and orbit angular momentum states $[^1S_0], ~[^3S_1], ~[^1P_1]$, and $[^3P_J]$ ($J=0,1,2$).
Here all the intermediated Fock states are in color singlet model (CSM).
The differential cross sections can be factored into the short-distance coefficients and the long-distance matrix elements \cite{nrqcd1,nrqcd2},
\begin{equation}
d\sigma(H_1,H_2)= d\hat\sigma((Q^\prime\bar{Q})[n]+ (Q\bar{Q^\prime})[n^\prime]) {\langle{\cal O}[n] \rangle} {\langle{\cal O}[n^\prime] \rangle}.
\label{dsigma}
\end{equation}
The short-distance coefficients $\hat\sigma$ describe the short-distance production of two Fock states $(Q^\prime\bar{Q})[n]$ and $(Q\bar{Q^\prime})[n^\prime]$ ($Q,~Q^\prime=c$ or $b$ quarks) with the spin and orbit angular momentum states $[n]$ and $[n^\prime]$, respectively.
In CSM, the non-perturbative long-distance matrix elements ${\langle{\cal O}[n^{(\prime)}] \rangle}$ describe the hadronization of the Fock state $(Q\bar{Q^\prime})[n^{(\prime)}]$ into the physical heavy quarkonia $H_{1(2)}$ in the same quantum states $[n^{(\prime)}]$.

The short-distance differential cross section $d\hat\sigma$ are perturbatively calculable,
\begin{eqnarray}
d\hat\sigma=\frac{1}{4\sqrt{(p_1\cdot p_2)^2-m^4_{e}}} \overline{\sum}  \big|{\cal M}([n],[n^\prime])\big|^{2} d\Phi_2,
\label{sd-sigma}
\end{eqnarray}
where $\overline{\sum}$ stands for the average over the spin of the initial electron and positron, and sum over the color and spin of the two Fock states.
In the $e^-e^+$ center-of-momentum (CM) frame, the two-body phase space can be simplified as
\begin{eqnarray}
    d{\Phi_2} &=& (2\pi)^4 \delta^{4}\left(p_1+p_2 - \sum_{f=1}^2 q_{f}\right)\prod_{f=1}^2 \frac{d^3{\vec{q}_f}}{(2\pi)^3 2q_f^0} \nonumber \\
    &=& \frac{{\mid\vec{q}_1}\mid}{8\pi\sqrt{s}}d(cos\theta).
    \label{dPhi-2}
\end{eqnarray}
Here, $s=(p_1+p_2)^2$ stands for the squared CM energy.
The magnitude of the 3-momentum ${|\vec{q}_1|}=\sqrt{\lambda[s, M^2_{1}, M^2_{2}] }/2\sqrt{s}$, where $M_{1(2)}$ are the masses of two heavy quarkonia $H_{1(2)}$ and $\lambda[a, b, c]= (a - b - c)^2 - 4 b c$.
The $\theta$ is the angle between the momentum $\vec{p_1}$ of the electron and the momentum $\vec{q_1}$ of the quarkonium $H_1[n]$.

The hard scattering amplitude ${\cal M}([n],[n^\prime])$ in Eq.~(\ref{sd-sigma}) can be read directly from the Feynman diagrams in Fig.~\ref{feyn1},
\begin{eqnarray}
i{\cal M}([n],[n^\prime])=  \sum_{j=1}^4 \bar{v}_{l^\prime} (p_2) {\cal V}^\mu u_l (p_1)  {\cal D}_{\mu\nu} {\cal A}^\nu_j,
\label{amplitude}
\end{eqnarray}
where $j$ represents the number of Feynman diagrams, and $l^{(\prime)}$ are the spins of the initial leptons.
The vertex $\cal{V}^\mu$ and the propagator $\cal{D_{\mu\nu}}$ for the virtual photon and $Z^0$ propagated processes have different forms,
\begin{eqnarray}
    \cal{V}^\mu &=&
       \alpha( -i e e_Q\gamma^\mu ) +
       \beta \frac{-i g} {4 cos\theta_W} \gamma^\mu (1-4 e_Q sin^2 \theta_W-\gamma^5), \nonumber \\
    \cal{D_{\mu\nu}} &=&
    \alpha \frac{-i g_{\mu\nu}}{p^2} +
     \beta   \frac{-i g_{\mu\nu}}{p^2-m^2_Z+i m_Z \Gamma_Z}.
\end{eqnarray}
In which, $\alpha=1, \beta=0$ is for virtual photon propagated processes, and $\alpha=0, \beta=1$ is for $Z^0$ propagated processes.
$e$ is the unit of the electric charge, $e_Q=1,2/3,-1/3$ for the leptons, $c$ quark, and $b$ quark, respectively. $g$ is the weak interaction coupling constant, and $\theta_W$ represents the Weinberg angle.
$p=p_1+p_2$ is the 4-momentum of the propagator.
$m_Z$ and $\Gamma_Z$ are the mass and the total decay width of $Z^0$ boson.

The explicit expressions of the Dirac $\gamma$ matrix chains ${\cal A}^\nu_j~(j=1,2,3,4)$ in Eq. (\ref{amplitude}) for the case that both two quarkonia or $B_c$'s are the $S$-wave states ($[^1S_0]$ , $[^3S_1]$) can be formulated as
\begin{widetext}
\begin{eqnarray}
{\cal A}^{\nu (S,L=0)}_1 &=& i  Tr \left[\Pi^{(S,L=0)}_{q_1} \gamma^\sigma  \frac{(\slashed{q}_1+ \slashed{q}_{21})+m_{Q_1}} {[(q_1 + q_{21})^2-m^2_{Q_1}] (q_{12} + q_{21})^2} {\cal V}^{\nu} \Pi^{(S,L=0)}_{q_2} \gamma^\sigma \right], \nonumber\\
{\cal A}^{\nu (S,L=0)}_2 &=& i  Tr \left[\Pi^{(S,L=0)}_{q_1} {\cal V}^{\nu}  \frac{-(\slashed{q}_2+ \slashed{q}_{12})+m_{{Q}_2}} {[(q_2 + q_{12})^2-m^2_{{Q}_2}] (q_{12} + q_{21})^2} \gamma^\sigma \Pi^{(S,L=0)}_{q_2} \gamma^\sigma \right], \nonumber\\
{\cal A}^{\nu (S,L=0)}_3 &=& i  Tr \left[\Pi^{(S,L=0)}_{q_1} \gamma^\sigma \Pi^{(S,L=0)}_{q_2} {\cal V}^{\nu} \frac{-(\slashed{q}_1+ \slashed{q}_{22})+m_{{Q}_1^\prime}} {[(q_1 + q_{22})^2-m^2_{{Q}_1^\prime}] (q_{11} + q_{22})^2} \gamma^\sigma \right],\nonumber\\
{\cal A}^{\nu (S,L=0)}_4 &=& i  Tr \left[\Pi^{(S,L=0)}_{q_1} \gamma^\sigma \Pi^{(S,L=0)}_{q_2} \gamma^\sigma  \frac{(\slashed{q}_2+ \slashed{q}_{11})+m_{Q_2^\prime}} {[(q_2 + q_{11})^2-m^2_{Q_2^\prime}] (q_{11} + q_{22})^2} {\cal V}^{\nu}  \right].
\label{HadAmp-S}
\end{eqnarray}
\end{widetext}
Here, $S$ and $L$ are the spin and the orbit angular momentum of the Fock states $[n^{(\prime)}]$. $q_{11}=\frac{m_{Q_1}}{M_{1}}{q_1}+q$ and $q_{12}=\frac{m_{Q_1^\prime}}{M_{1}}{q_1}-q$ are the momenta of the two constituent heavy quarks of the quarkonium $H_1[n](q_1)$
with $q$ being the relative momentum between them and $M_{1}=m_{Q_1}+m_{{Q}_1^\prime}$.
$q_{21}=\frac{m_{Q_2^\prime}}{M_{2}}{q_2}+q^\prime$ and $q_{22}=\frac{m_{{Q}_2}}{M_{2}}{q_2}-q^\prime$ are the momenta of the two quarks of the quarkonium $H_2[n^\prime](q_2)$ with $q^\prime$ being the relative momentum between them and $M_{2}=m_{Q_2^\prime}+m_{{Q}_2}$.
The two projectors $\Pi^{(S,L=0)}_{q_k}$ ($k=1, 2$) in Eq. (\ref{HadAmp-S}) have the following forms
\begin{eqnarray}
\Pi^{(S,L=0)}_{q_k} &=& \epsilon_a(q_k) \frac{-\sqrt{M_{k}}}{4 m_{Q_k}m_{Q^\prime_k}}(\slashed{q}_{k2}- m_{Q^\prime_k}) \gamma^a (\slashed{q}_{k1} + m_{Q_k})\otimes\frac{\delta_{ij}} {\sqrt{N_c}}.
\end{eqnarray}
For spin-singlet state $[^1S_0]$, $\epsilon_a(q_k)=1$ and $\gamma^a=\gamma^5$; for spin-triplet state $[^3S_1]$, $\epsilon_a(q_k)=\varepsilon_{\alpha}(q_k)$ is its polarization vector and $\gamma^a=\gamma^\alpha$ is the Dirac matrix with Lorentz index $\alpha$.
For the $S$-wave states, the relative momentum $q$ and $q^{\prime}$ are set to zero directly.
In CSM, $\delta_{ij}/\sqrt{N_c}$ is the color operator for color-singlet projector of the heavy quarkonium with $N_c=3$.

We shall consider the case that one of two quarkonia or $B_c$ mesons $H_2[n^\prime](q_2)$ is a $P$-wave state ($[^1P_1]$, $[^3P_J~(J=0,1,2)]$).
The expressions of ${\cal A}^\nu_j~(j=1,2,3,4)$ for this case can be expressed in those for both $S$-wave case,
\begin{eqnarray}
{\cal A}^{\nu (S=0, L=1)}_j &=&  \varepsilon_\beta(q_2) \left. \frac{d}{d q^\prime_\beta} {\cal A}^{\nu (S=0, L=0)}_j \right|_{q=0,q^\prime=0}, \nonumber\\
{\cal A}^{\nu (S=1, L=1)}_j &=&  \varepsilon^J_{\alpha \beta} (q_2)\left. \frac{d}{d q^\prime_\beta} {\cal A}^{\nu (S=1, L=0)}_j \right|_{q=0,q^\prime=0}.
\label{HadAmp-P}
\end{eqnarray}
In which, $\varepsilon_\beta(q_2)$ is the polarization vector of the $[^1P_1]$ state, and $\varepsilon^{J}_{\alpha\beta}(q_2)$ is the polarization tensor for $[^3P_J]$ states with $J=0,1,2$.
The derivatives over the relative momentum $q^\prime_\beta$ of the quarkonium or $B_c$ meson $H_2[n^\prime]$ will give complicated and lengthy amplitudes.
The relative momentum $q^\prime$ is set to zero after taking the derivatives.

To get compact analytical expression of the complicated $P$-wave channels and also to improve the efficiency of numerical evaluation, we adopt the ``improved trace technology" to simplify the amplitudes ${\cal M}([n],[n^\prime])$ at the amplitude level before evaluating the polarization sum.
To shorten this manuscript, we don't present the prescription. For detailed techniques and examples, one can refer to the literatures \cite{cjx,lxz,Yang:2011ps,wbc1,Liao:2012rh,Liao:2021ifc}.

When manipulating the squared amplitudes $|{\cal M}([n],[n^\prime])|^{2}$ in Eq.~(\ref{sd-sigma}), we need to sum over the polarization vectors of the heavy quarkonia.
For the spin-triplet state $[^3S_1]$ or the spin-singlet state $[^1P_1]$ with 4-momentum $p$, the polarization sum is given by \cite{nrqcd2}
\begin{eqnarray}
\sum_{J_z}\varepsilon_{\alpha} \varepsilon_{\alpha^\prime} = \Pi_{\alpha\alpha^\prime} \equiv -g_{\alpha \alpha^\prime}+\frac{p_{\alpha} p_{\alpha^\prime}}{p^2},
\end{eqnarray}
where $J_z=S_z$ or $L_z$ for $[^3S_1]$ and $[^1P_1]$ states, respectively. In the case of $[^3P_J]$ states, the polarization sum should be performed by the selection of appropriate total angular momentum quantum number $J$.
The sum over polarization tensors  is given by \cite{nrqcd2}
\begin{eqnarray}
\varepsilon^{0}_{\alpha\beta} \varepsilon^{0*}_{\alpha^\prime\beta^\prime} &=& \frac{1}{3} \Pi_{\alpha\beta}\Pi_{\alpha^\prime\beta^\prime}, \nonumber\\
\sum_{J_z}\varepsilon^{1}_{\alpha\beta} \varepsilon^{1*}_{\alpha^\prime\beta^\prime} &=& \frac{1}{2}
(\Pi_{\alpha\alpha^\prime}\Pi_{\beta\beta^\prime}- \Pi_{\alpha\beta^\prime}\Pi_{\alpha^\prime\beta}),\nonumber\\
\sum_{J_z}\varepsilon^{2}_{\alpha\beta} \varepsilon^{2*}_{\alpha^\prime\beta^\prime} &=& \frac{1}{2}
(\Pi_{\alpha\alpha^\prime}\Pi_{\beta\beta^\prime}+ \Pi_{\alpha\beta^\prime}\Pi_{\alpha^\prime\beta})-\frac{1}{3} \Pi_{\alpha\beta}\Pi_{\alpha^\prime\beta^\prime}, \nonumber
\end{eqnarray}
for total angular momentum $J=0,1,2$, respectively.

At last we discuss the hadronization of the Fock states into the physical heavy quarkonia, which is described by the nonperturbative color-singlet matrix element $\langle{\cal O}([n^{(\prime)}])\rangle$ in Eq.~(\ref{dsigma}).
The heavy-quark spin symmetry provides relations between matrix elements for the various spin states \cite{nrqcd1},
\begin{eqnarray}
\langle{\cal O}([^3S_1]) \rangle &=& \langle{\cal O}([^1S_0]) \rangle[1+O(v^2)],\nonumber\\
\langle{\cal O}([^3P_J]) \rangle &=& \langle{\cal O}([^1P_0]) \rangle[1+O(v^2)].
\end{eqnarray}
Here, $v$ is the relative velocity between the constituent heavy quark and antiquark in the quarkonium or $B_c$ rest frame.
Further, vacuum-saturation approximation together with the heavy-quark spin symmetry can be used to express all
the color-singlet matrix elements in terms of the radial wave functions at the origin $R_{S}(0)$ for $S$-wave states and their first derivatives at the origin $R^\prime_{P}(0)$ for $P$-wave states \cite{nrqcd1},
\begin{eqnarray}
\langle{\cal O}([^1S_0])\rangle &=& \frac{1}{4\pi} \big|R_{S}(0)\big|^2 [1+O(v^4)],\nonumber\\
\langle{\cal O}([^3S_1])\rangle &=& \frac{1}{4\pi} \big|R_{S}(0)\big|^2 [1+O(v^4)],\nonumber\\
\langle{\cal O}([^1P_1])\rangle &=& \frac{3}{4\pi} \big|R^\prime_{P}(0)\big|^2 [1+O(v^2)],\nonumber\\
\langle{\cal O}([^3P_J])\rangle &=& \frac{3}{4\pi} \big|R^\prime_{P}(0)\big|^2 [1+O(v^2)].
\end{eqnarray}
Since the relativistic corrections are at order $v^2$ or evern higher in NRQCD, we can adopt the same values of matrix elements for both the spin-singlet and spin-triplet states.
Moreover, the radial wave functions (or their derivatives) of the heavy quarkonia or $B_c$ mesons at the origin can be extracted from the decays of the quarkonia or $B_c$'s in experiments, and be calculated within the potential models \cite{wgs,pot2,Richardson:1978bt,IO,Ikhdair:2003ry,Chen:1992fq,Eichten:1978tg,Eichten:1979ms,Eichten:1980mw,Eichten:1995ch} or potential NRQCD \cite{Brambilla:1999xf} or lattice QCD \cite{Bodwin:1996tg}.

%%%%%%%%%%%%%%%%
%%%%%%%%%%%%%%%%
\section{Phenomenology}
%%%%%%%%%%%%%%%%
%%%%%%%%%%%%%%%%
\subsection{Input parameters}
%%%%%%%%%%%%%%%%%%
\begin{table}
\caption{The masses (units: GeV) of the constituent quarks, and the squared radial wave functions at the origin $|R_{S}(0)|^2$ (units: GeV$^3$) and their first derivatives at the origin $|R^\prime_{P}(0)|^2$ (units: GeV$^5$) for the charmonia, bottomonia and $B_c$ mesons within the BT-potential model~\cite{lx}.
The uncertainties of radial wave functions (or their first derivatives) at the origin are caused by the corresponding varying quark masses.}
\begin{tabular}{|c|c|c|c|c|}
\hline
&\multicolumn{2}{|c|}{$S$-wave} & \multicolumn{2}{|c|}{$P$-wave}\\
\hline\hline
\multirow{2}{*}{Chamonia} & $m_c$ & $|R_{(c\bar{c})[S]}(0)|^2$ & $m_c$ & $|R^\prime_{(c\bar{c})[P]}(0)|^2$\\
\cline{2-5}
& 1.48$\pm$0.1 & $2.458^{+0.227}_{-0.327}$&1.75$\pm$0.1 & $0.322^{+0.077}_{-0.068}$\\
\hline\hline
\multirow{2}{*}{Bottomonia} & $m_b$ & $|R_{(b\bar{b})[S]}(0)|^2$&$m_b$ & $|R^\prime_{(b\bar{b})[P]}(0)|^2$\\
\cline{2-5}
& 4.71$\pm$0.2 & $16.12^{+1.28}_{-1.23}$&4.94$\pm$0.2 & $5.874^{+0.728}_{-0.675}$\\
\hline\hline
\multirow{2}{*}{$B_c$ mesons} & $m_c,~m_b$ & $|R_{c\bar{b}[S]}(0)|^2$ & $m_c,~m_b$ & $|R^\prime_{c\bar{b}[P]}(0)|^2$\\
\cline{2-5}
& 1.45$\pm$0.1,~~4.85$\pm$0.2 & ~~$3.848^{+0.474}_{-0.453}$ & 1.75$\pm$0.1,~~4.93$\pm$0.2 & ~~$0.518^{+0.123}_{-0.105}$\\
\hline
\end{tabular}
\label{M&R}
\end{table}
%%%%%%%%%%%%%%%%%%%%%

For the numerical analysis, masses of the constituent $c$ and $b$ quarks for heavy quarkonia and $B_c$ mesons are displayed in Table \ref{M&R}.
The mass of a heavy quarkonium is always set to be the sum of the masses of its constituent quarks $M_{i}=m_{Q_i}+m_{Q^\prime_i}~(i=1,2)$, which ensures the gauge invariance of the hard scattering amplitude under the NRQCD framework.
Thus we adopt different constituent quark masses for the $S$- and $P$-wave heavy quarkonia and $B_c$ mesons.
In our previous work \cite{lx}, the radial wave functions at the origin $|R_{S}(0)|^2$ and the first derivatives of radial wave functions at the origin $|R^\prime_{P}(0)|^2$ for various heavy quarkonia and $B_c$ mesons have been calculated under five different potential models.
In this work, we adopt the results of the Buchm\"{u}ller and Tye potential model (BT-potential) \cite{lx,wgs,pot2}, which are also presented in Table \ref{M&R}.
The uncertainties of radial wave functions (or their derivatives) at the origin are caused by the corresponding varying quark masses, which would be taken into consideration when we discuss the uncertainties caused by varying quark masses in Sec. \ref{uncertainty}.
The LO running strong coupling constant $\alpha_s$ is adopted, which leads to $\alpha_s(2m_c)=0.26$ for charmonia and $B_c$ mesons, and $\alpha_s(2m_b)=0.18$ for bottomonia.
Other parameters have the following values \cite{pdg}: the Fermi constant $G_F=\frac{\sqrt{2}g^2}{8m_W^2}=1.16639 \times 10^{-5}$ GeV$^{-2}$ with $m_W= 80.399$ GeV, the fine structure constant $\alpha=e^2/4\pi=1/130.9$, the Weinberg angle $\theta_W=\arcsin\sqrt{0.23119}$, and the $Z^0$ boson mass $m_Z =91.1876$ GeV and its total decay width $\Gamma_{Z^0}=2.4952$ GeV.

%%%%%%%%%%%%%%%%%%%%%%%%%%%%%%%%
\subsection{Total and differential cross sections}
\label{production}
%%%%%%%%%%%%%%%%%%%%%%%%%%%%%%%%
%%%%%%%%%%%%%%%%%%%%%%
\begin{table}
\caption{Cross sections (units:~$fb$) for the production of the double heavy quarkonia and $B_c$ mesons in $e^+e^-$ annihilation at $\sqrt{s}=m_Z$ within the BT-potential model~\cite{lx}. The subscripts $\gamma^*$ and $Z^0$ are for virtual photon and $Z^0$ boson propagated processes, respectively. The top three channels in each column are in bold face.}
\begin{tabular}{|c|c|c|c|c|c|}
\hline
\multicolumn{2}{|c|}{Charmonia} & \multicolumn{2}{|c|}{Bottomonia} & \multicolumn{2}{|c|}{$B_c$ mesons}\\
\hline\hline
$\sigma{(\eta_c+J/\psi)}_{\gamma^*}$ &~3.269~$\times 10^{-6}$~&$\sigma{(\eta_b+\Upsilon)}_{\gamma^*}$ &~1.588~$\times 10^{-5}$~&$\sigma{(\eta_{bc}^++B_c^{*-})}_{\gamma^*}$&~5.711~$\times 10^{-5}$~\\
\hline
$\sigma{(\eta_c+h_c)}_{\gamma^*}$ &~1.051~$\times 10^{-5}$~&$\sigma{(\eta_b+h_{b})}_{\gamma^*}$ &~2.021~$\times 10^{-6}$~&$\sigma{(\eta_{bc}^++h_{bc}^-)}_{\gamma^*}$&~2.288~$\times 10^{-6}$~\\
\hline
$\sigma{(J/\psi+h_c)}_{\gamma^*}$ &~8.821~$\times 10^{-8}$~&$\sigma{(\Upsilon+h_{b})}_{\gamma^*}$ &~1.618~$\times 10^{-7}$~&$\sigma{(B_c^{*+}+h_{bc}^-)}_{\gamma^*}$&~1.746~$\times 10^{-6}$~\\
\hline
$\sigma{(J/\psi+\chi_{c0})}_{\gamma^*}$ &~1.084~$\times 10^{-5}$~&$\sigma{(\Upsilon+\chi_{b0})}_{\gamma^*}$ &~3.747~$\times 10^{-6}$~&$\sigma{(B_c^{*+}+\chi_{bc0}^-)}_{\gamma^*}$&~1.039~$\times 10^{-5}$~\\
\hline
$\sigma{(J/\psi+\chi_{c1})}_{\gamma^*}$ &~2.142~$\times 10^{-5}$~&$\sigma{(\Upsilon+\chi_{b1})}_{\gamma^*}$ &~8.293~$\times 10^{-6}$~&$\sigma{(B_c^{*+}+\chi_{bc1}^-)}_{\gamma^*}$&~8.443~$\times 10^{-5}$~\\
\hline
$\sigma{(J/\psi+\chi_{c2})}_{\gamma^*}$ &~4.264~$\times 10^{-5}$~&$\sigma{(\Upsilon+\chi_{b2})}_{\gamma^*}$ &~1.509~$\times 10^{-5}$~&$\sigma{(B_c^{*+}+\chi_{bc2}^-)}_{\gamma^*}$&~2.816~$\times 10^{-4}$~\\
\hline \hline
$\sigma{(\eta_c+J/\psi)}_{Z^0}$ &~1.895~$\times 10^{-4}$~&\bm{$\sigma{(\eta_b+\Upsilon)}_{Z^0}$}&\bm{$~4.286~\times 10^{-2}$}~&\bm{$\sigma{(\eta_{bc}^++B_c^{*-})}_{Z^0}$}&\bm{$~0.6346~$}\\
\hline
$\sigma{(J/\psi+J/\psi)}_{Z^0}$ &~6.416~$\times 10^{-4}$~&\bm{$\sigma{(\Upsilon+\Upsilon)}_{Z^0}$}&\bm{$~1.199~\times 10^{-2}$}~&\bm{$\sigma{(B_c^{*+}+B_c^{*-})}_{Z^0}$}&\bm{$~1.150~$}\\
\hline
$\sigma{(\eta_c+h_c)}_{Z^0}$ &~6.076~$\times 10^{-4}$~&$\sigma{(\eta_b+h_{b})}_{Z^0}$&~5.327~$\times 10^{-3}$~&$\sigma{(\eta_{bc}^++h_{bc}^-)}_{Z^0}$&~4.810~$\times 10^{-3}$~\\
\hline
\bm{$\sigma{(J/\psi+h_c)}_{Z^0}$} &~\bm{$4.540~\times 10^{-3}$}~&$\sigma{(\Upsilon+h_{b})}_{Z^0}$&~6.146~$\times 10^{-3}$~&$\sigma{(B_c^{*+}+h_{bc}^-)}_{Z^0}$&~1.177~$\times 10^{-2}$~\\
\hline
\bm{$\sigma{(\eta_c+\chi_{c0})}_{Z^0}$} &~\bm{$1.370~\times 10^{-3}$}~&$\sigma{(\eta_b+\chi_{b0})}_{Z^0}$&~9.215~$\times 10^{-4}$~&$\sigma{(\eta_{bc}^++\chi_{bc0}^-)}_{Z^0}$&~1.527~$\times 10^{-2}$~\\
\hline
$\sigma{(\eta_c+\chi_{c1})}_{Z^0}$ &~1.887~$\times 10^{-4}$~&$\sigma{(\eta_b+\chi_{b1})}_{Z^0}$&~1.431~$\times 10^{-3}$~&$\sigma{(\eta_{bc}^++\chi_{bc1}^-)}_{Z^0}$&~6.162~$\times 10^{-3}$~\\
\hline
\bm{$\sigma{(\eta_c+\chi_{c2})}_{Z^0}$} &~\bm{$2.842~\times 10^{-3}$}~&$\sigma{(\eta_b+\chi_{b2})}_{Z^0}$&~2.815~$\times 10^{-3}$~&$\sigma{(\eta_{bc}^++\chi_{bc2}^-)}_{Z^0}$&~3.093~$\times 10^{-3}$~\\
\hline
$\sigma{(J/\psi+\chi_{c0})}_{Z^0}$ &~4.403~$\times 10^{-4}$~&$\sigma{(\Upsilon+\chi_{b0})}_{Z^0}$&~6.837~$\times 10^{-3}$~&$\sigma{(B_c^{*+}+\chi_{bc0}^-)}_{Z^0}$&~0.4542~\\
\hline
$\sigma{(J/\psi+\chi_{c1})}_{Z^0}$ &~1.762~$\times 10^{-4}$~&$\sigma{(\Upsilon+\chi_{b1})}_{Z^0}$&~1.175~$\times 10^{-2}$~&$\sigma{(B_c^{*+}+\chi_{bc1}^-)}_{Z^0}$&~0.2701~\\
\hline
$\sigma{(J/\psi+\chi_{c2})}_{Z^0}$ &~5.751~$\times 10^{-4}$~&\bm{$\sigma{(\Upsilon+\chi_{b2})}_{Z^0}$}& \bm{$~2.306~\times 10^{-2}$}~&\bm{$\sigma{(B_c^{*+}+\chi_{bc2}^-)}_{Z^0}$}&\bm{$~1.133~$}\\
\hline
\end{tabular}
\label{tabrpa}
\end{table}
%%%%%%%%%%%%%%%%%%%%%%

The total cross sections for the production of double heavy quarkonia and $B_c$ mesons via $e^- e^+ \to \gamma^*/Z^0 \to H_1[n] + H_2[n^\prime]$ at CM energy $\sqrt{s}=91.1876$ GeV are listed in Table \ref{tabrpa}.
From this table, one could draw a conclusion that the contributions from virtual photon propagated processes are negligible in comparison with $Z^0$ propagated processes at future super $Z^0$ factory.
The top three $Z^0$ propagated channels for the production of double charmonia, double bottomonia, and double $B_c$ mesons are
\begin{eqnarray}
\sigma{(J/\psi+h_c)}_{Z^0} &>&
\sigma{(\eta_c+\chi_{c2})}_{Z^0}>
\sigma{(\eta_c+\chi_{c0})}_{Z^0} > \dots, \nonumber\\
\sigma{(\eta_b+\Upsilon)}_{Z^0} &>& \sigma{(\Upsilon+\chi_{b2})}_{Z^0} >
\sigma{(\Upsilon+\Upsilon)}_{Z^0} > \dots, \nonumber\\
\sigma{(B_c^{*+}+B_c^{*-})}_{Z^0} &>& \sigma{(B_c^{*+}+\chi_{bc2}^-)}_{Z^0} >
\sigma{(\eta_{bc}^++B_c^{*-})}_{Z^0} > \dots
\end{eqnarray}
In Ref. \cite{gxz1}, Chen {\it et. al.} calculate the cross sections for the production of the double charmonia for $S$- and $P$-wave states in CSM. If the same input parameters are adopted, our estimations in Table \ref{tabrpa} are consistent with theirs.

For the $Z$ factory operation mode at CEPC, the designed integrated luminosity with two interaction point and in two years is $16~ab^{-1}$ \cite{CEPCStudyGroup:2018ghi}.
Then we can estimate the events of the production of double heavy quarkonia and $B_c$ mesons.
The top three channels for double charmonia are
73 $J/\psi+h_c$ events,
45 $\eta_c+\chi_{c2}$ events, and
22 $\eta_c+\chi_{c0}$ events.
The top three channels for double bottomonia are
686 $\eta_b+\Upsilon$ events,
369 $\Upsilon+\chi_{b2}$ events, and
192 $\Upsilon+\Upsilon$ events.
The top three channels for double $B_c$ mesons are
$1.84\times10^4$ $B_c^{*+}+B_c^{*-}$ events,
$1.81\times10^4$ $B_c^{*+}+\chi_{bc2}^-$ events, and
$1.02\times10^4$ $\eta_{bc}^++B_c^{*-}$ events.
We take the $J/\psi+h_c$ channel as an example to discuss its observation at CEPC. The $J/\psi$ event can be fully reconstructed through the leptonic decay $J/\psi \to l^+ l^-$ ($l=e,\mu$) whose branching fraction is about 12\% \cite{pdg}.
The $h_c$ can be reconstructed through the decay chain $h_c \to \eta_c+ \gamma \to (\text{hadronic decays})+\gamma$ with the $(50\pm9)$\% branching fraction of $h_c \to \eta_c + \gamma$ \cite{pdg} and about total 43\% branching fraction of the sixteen hadronic decays of $\eta_c$\footnote{The sixteen decay chains for the reconstruction of $\eta_c$ can be found in the talk ``Hadronic Transitions in $e^+e^-$ Collisions above 4 GeV'' by Jianming Bian (Minnesota U.) in the 11th International Workshop on Heavy Quarkonium (QWG2016) at \url{https://boss.ihep.ac.cn/~talks/index.php/Talks\_in\_2016}. And their explicit branching fractions can be found in PDG \cite{pdg}.}.
Thus with two years' data, we can obtain only 2 events of $J/\psi+h_c$ in the $Z$ factory mode at the CEPC.
Although the events of double bottomonia and double $B_c$ meosns are larger than the events of double charmonia by about one and two orders of magnitude respectively, the observations for the production of them are also not optimistic because of the bigger suppression factors from their decays.\footnote{The leptonic branching fraction of $\Upsilon \to l^+ l^-$ ($l=e,\mu,\tau$) are about 7.5\%. The reconstruction of excited bottomonia can be reconstructed through their radiative dacay into $\Upsilon$, so they have even bigger suppresion factors. The $B_c(^1S_0)$ can be reconstructed by the decay mode $B_c^+(^1S_0) \to J/\psi \pi^+$, but its branching fraction is only about 0.2\%. We don't observe the $B_c^*(^3S_1)$ and $P$-wave $B_c$ mesons yet.}
In one word, two years' operation time for $Z$ factory mode at the CEPC is not sufficient to make the study on the production of double heavy quarkonia and $B_c$ mesons.
And the observation of such decays at the CEPC with sizable sample might indicates the presence of new production mechanism, for example the contributions of color-octet Fock states might dominate.

The FCC-ee would run as a super $Z$ factory in its first four years with the designed integrated luminosity of $150~ab^{-1}$ \cite{Agapov:2022bhm}, which is about nine times of that at CEPC.
Then the events of the top three channels for double charmonia, double bottomonia and double $B_c$ mesons are about 9 times of those at the CEPC.
Taking the double charmonium channels as an example, we can obtain
681 $J/\psi+h_c$ events,
426 $\eta_c+\chi_{c2}$ events, and
206 $\eta_c+\chi_{c0}$ events.
For the $J/\psi+h_c$ channel, we might reconstruct about 18 signals at the detector after considering the suppressions from the branching fractions.
But if we further consider the reconstruction efficiency in experiments, the prospects are not optimistic.
Anyway, the $Z$ factory operation mode at future FCC-ee would be a better choice to study the production of double heavy quarkonia and $B_c$ mesons than at CEPC.
It is worth noting that the CMS collaboration at the LHC carry out the search for $Z$ boson decays into $J/\psi$ and $\Upsilon$ pairs very recently \cite{CMS:2022fsq}. No evidence is found in the channels of $Z\to J/\psi J/\psi$, $Z \to \Upsilon(1S) \Upsilon(1S)$ and $Z \to \Upsilon(nS) \Upsilon(mS)$, but the upper limits for the branching fractions at 95\% confidence level are presented.

%%%%%%%%%%%%%%%%%%%%
\begin{figure*}[htbp]
\centering
\includegraphics[width=0.45\textwidth]{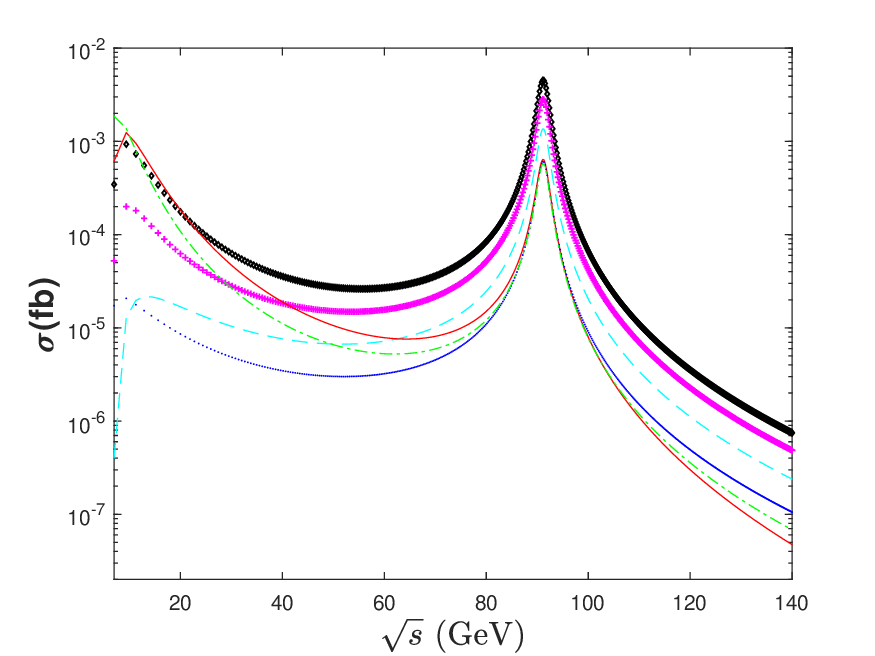}
\includegraphics[width=0.45\textwidth]{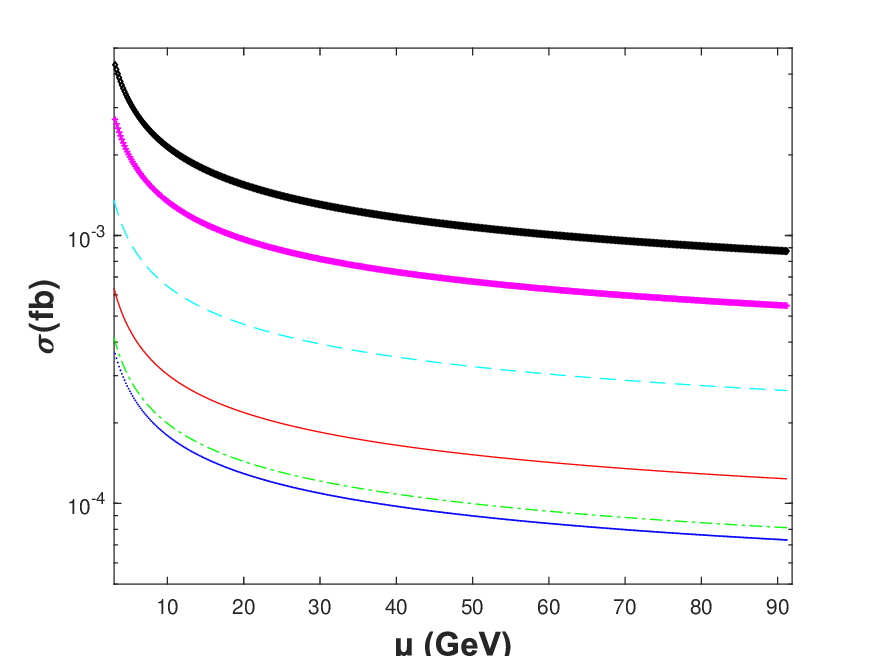}
\includegraphics[width=0.45\textwidth]{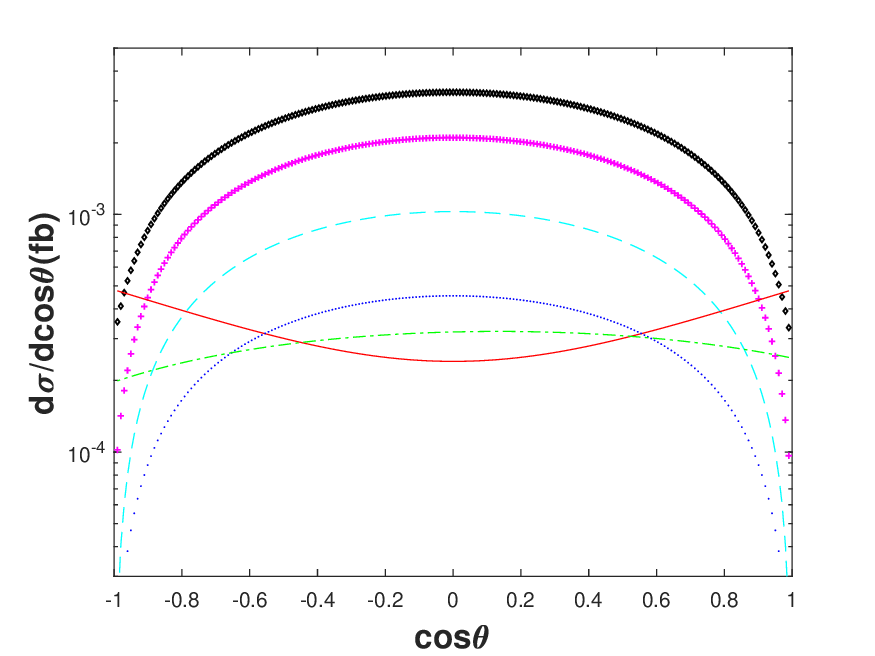}
\includegraphics[width=0.45\textwidth]{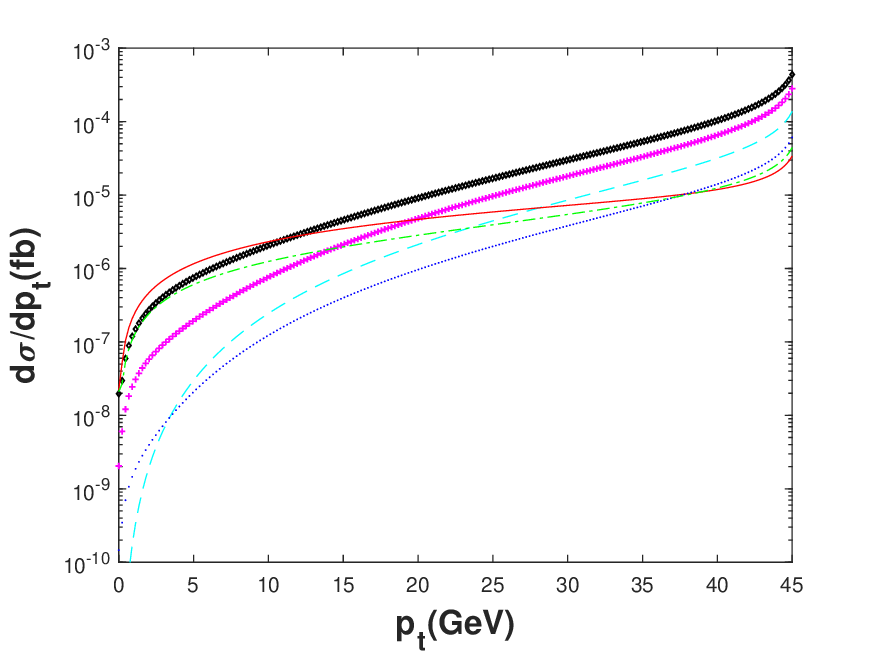}
\caption{
Total Cross sections versus the CM energy $\sqrt{s}$ and the renormalization scale $\mu$, the differential angle distributions d$\sigma$/dcos$\theta$, and the $p_t$ distributions d$\sigma$/d$p_t$ at $\sqrt{s}=m_Z$ for the double charmonium production.
The diamond black line, cross magenta line, dashed cyan line, solid red line, dotted blue line, and the dash-dotted green line are for the top six $Z^0$ propagated channels in Table \ref{tabrpa}:
$J/\psi+h_c$,
$\eta_c+\chi_{c2}$,
$\eta_c+\chi_{c0}$,
$J/\psi+J/\psi$,
$\eta_c+h_c$, and
$J/\psi+\chi_{c2}$, respectively.} \label{ccds}
\end{figure*}
%%%%%%%%%%%%%%%%%%%%

%%%%%%%%%%%%%%%%%%%%
\begin{figure*}[htbp]
\centering
\includegraphics[width=0.45\textwidth]{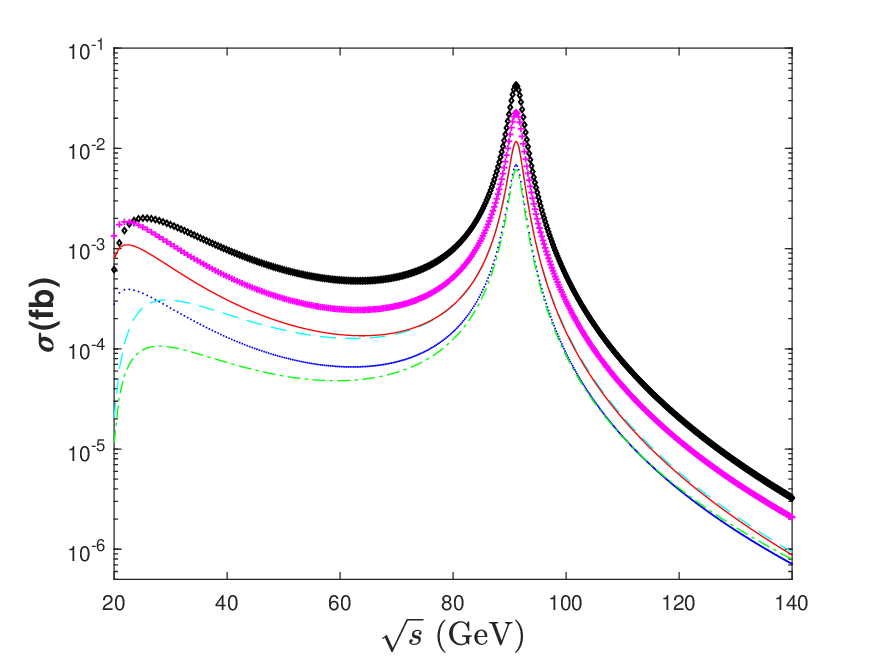}
\includegraphics[width=0.45\textwidth]{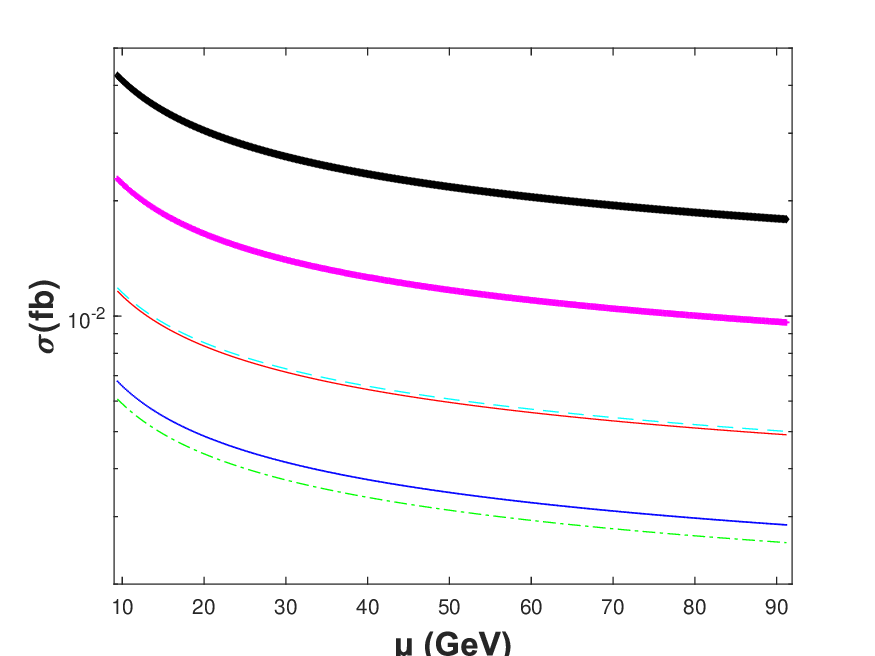}
\includegraphics[width=0.45\textwidth]{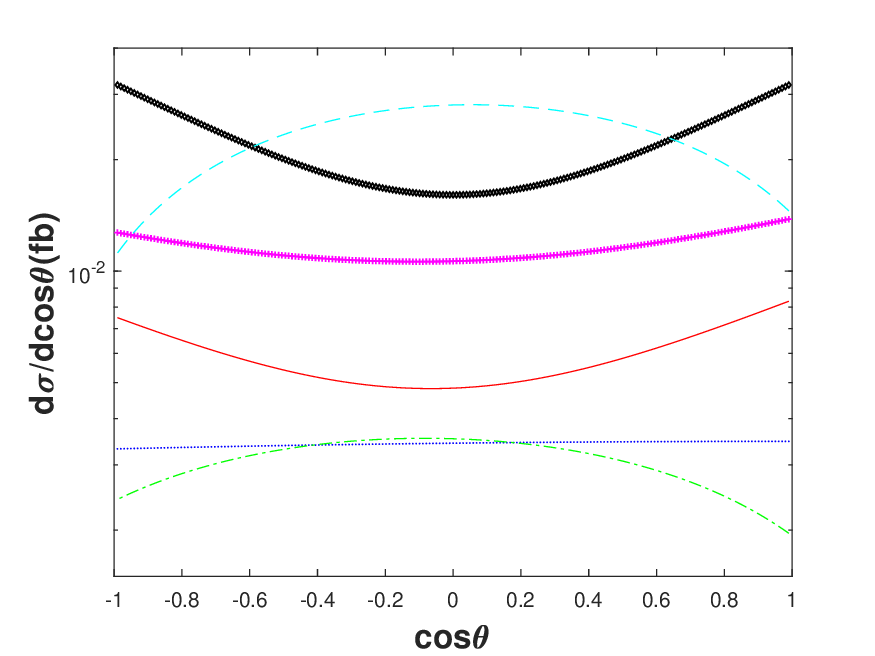}
\includegraphics[width=0.45\textwidth]{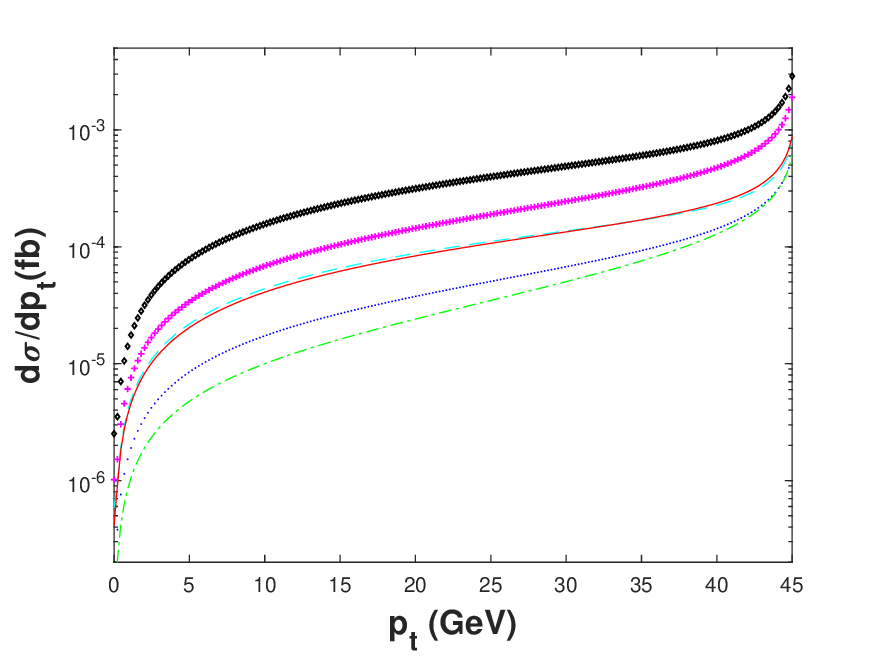}
\caption{
Total cross sections versus the CM energy $\sqrt{s}$ and the renormalization scale $\mu$, the differential angle distributions d$\sigma$/dcos$\theta$, and the $p_t$ distributions d$\sigma$/d$p_t$ and at $\sqrt{s}=m_Z$ for the double bottomonium production. The diamond black line, cross magenta line, dashed cyan line, solid red line, dotted blue line, and the dash-dotted green line are for the top six $Z^0$ propagated channels in Table \ref{tabrpa}:
$\eta_b+\Upsilon$,
$\Upsilon+\chi_{b2}$,
$\Upsilon+\Upsilon$,
$\Upsilon+\chi_{b1}$,
$\Upsilon+\chi_{b0}$,
and $\Upsilon+h_{b}$, respectively.} \label{bbds}
\end{figure*}
%%%%%%%%%%%%%%%%%%%%

%%%%%%%%%%%%%%%%%%%%
\begin{figure*}[htbp]
\centering
\includegraphics[width=0.45\textwidth]{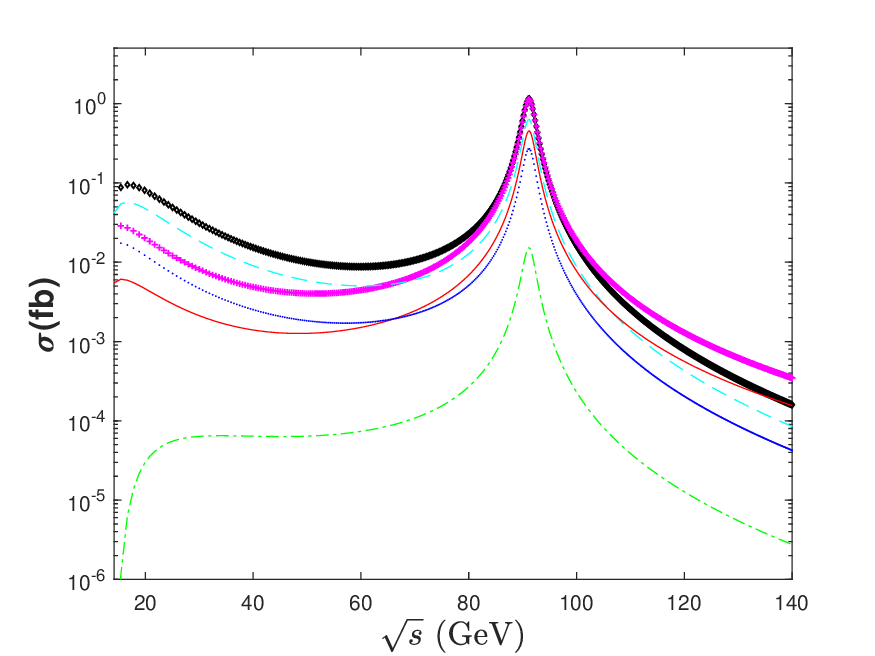}
\includegraphics[width=0.45\textwidth]{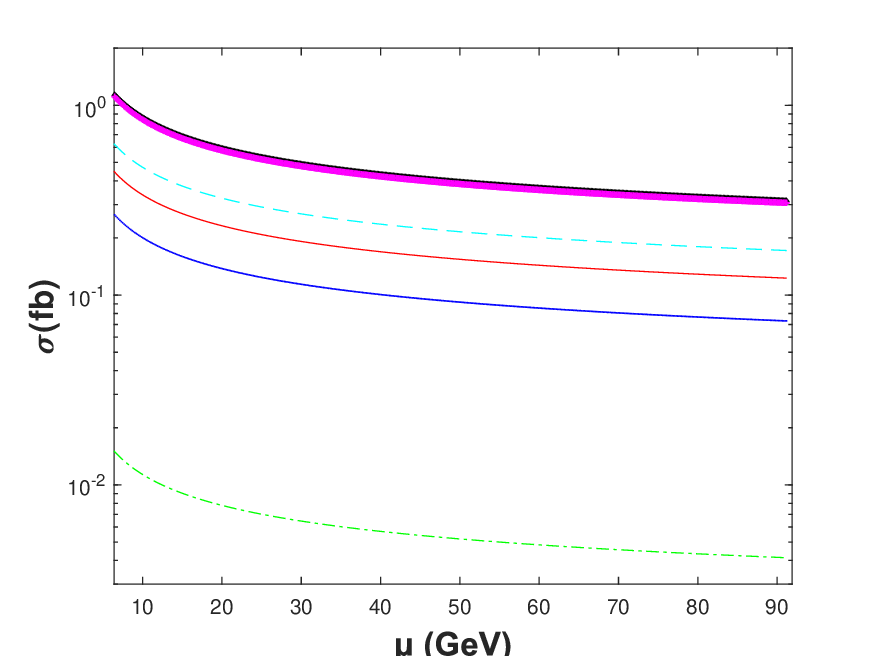}
\includegraphics[width=0.45\textwidth]{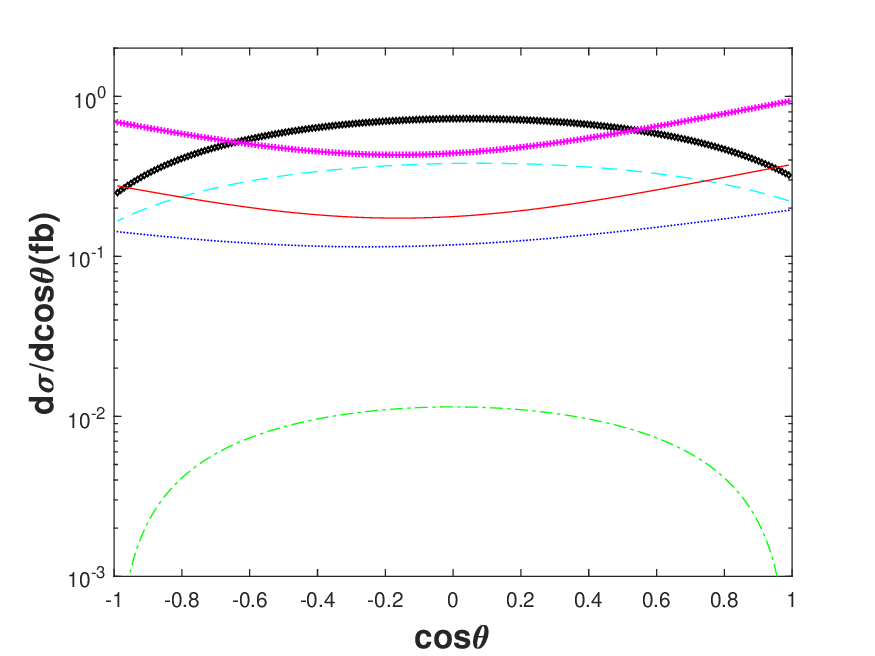}
\includegraphics[width=0.45\textwidth]{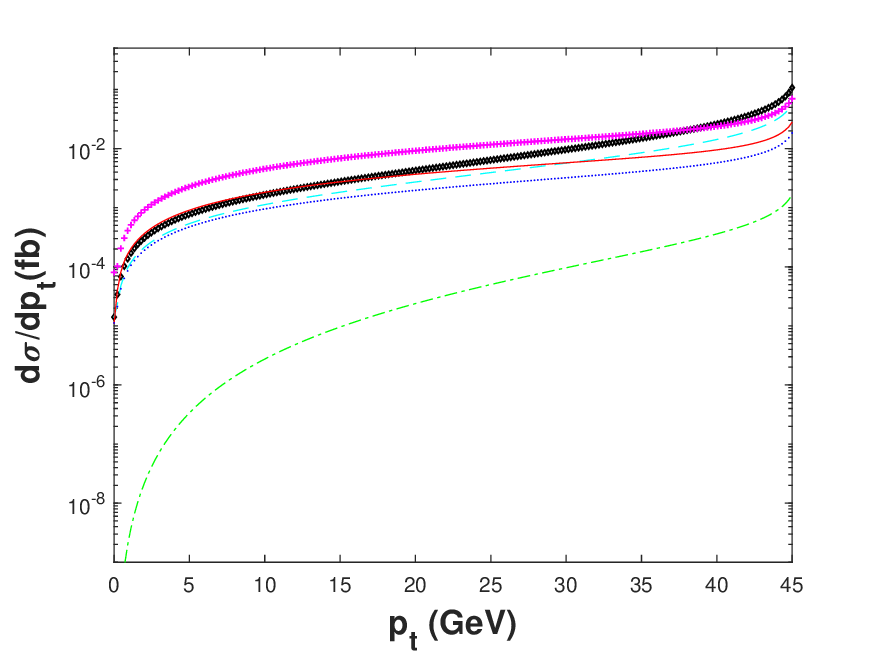}
\caption{
Total cross sections versus the CM energy $\sqrt{s}$ and the renormalization scale $\mu$, the differential angle distributions d$\sigma$/dcos$\theta$, and the $p_t$ distributions d$\sigma$/d$p_t$ at $\sqrt{s}=m_Z$ for the double $B_c$ production. The diamond black line, cross magenta line, dashed cyan line, solid red line, dotted blue line, and the dash-dotted green line are for the top six $Z^0$ propagated channels in Table \ref{tabrpa}:
$B_c^{*+}+B_c^{*-}$,
$B_c^{*+}+\chi_{bc2}^-$,
$\eta_{bc}^++B_c^{*-}$,
$B_c^{*+}+\chi_{bc0}^-$,
$B_c^{*+}+\chi_{bc1}^-$,
and $\eta_{bc}^++\chi_{bc0}^-$, respectively.} \label{bcds}
\end{figure*}
%%%%%%%%%%%%%%%%%%%%

In Figs. \ref{ccds}-\ref{bcds}, we display the total cross sections $\sigma(\sqrt{s})$ as functions of the CM energy $\sqrt{s}$, the total cross sections $\sigma(\mu)$ as functions of the renormalization scale $\mu$ in running strong coupling constant $\alpha_s(\mu)$, the differential angle distributions d$\sigma$/dcos$\theta$ and the $p_t$ distributions d$\sigma$/d$p_t$ at $\sqrt{s}=m_Z$ for the double charmonia, the double bottomonia and the double $B_c$ mesons, respectively.
In those figures, we only exhibit the curves of top six $Z^0$ propagated processes in Table \ref{tabrpa}.
Note that the figure legends in those three figures are different.

In the plots for total cross sections with running CM energy $\sqrt{s}$ in Figs. \ref{ccds}-\ref{bcds}, one can observe the dramatic peaks of cross section caused by the $Z^0$ resonance clearly.
In the future super $Z$ factory, the beam energy can be controlled with a very high accuracy.
The natural beam energy spread scales as $\sigma_E/E$ with the energy resolution $\sigma_E$ at the energy $E$.
At FCC-ee, the spread is 27 MeV \cite{FCC:2018evy}.
At CEPC, the expected accuracy is the order of 100 keV for the Z factory operation \cite{CEPCStudyGroup:2018ghi}.
However, the initial state radiation (ISR) at an electron-positron collider may make the CM energy $\sqrt{s}$ deviates from the $Z^0$ mass pole heavily.
The influence of ISR on the production rates can be identified as the deviations of $\sqrt{s}$ from $m_Z$.
In Table \ref{sqrtsdeviation}, we display the top six cross sections for the double heavy quarkonia and $B_c$ mesons with the deviations of both 1\%$m_Z$ and 3\%$m_Z$.
It is found that the total cross sections decrease by about 30\%$\sim$40\% when CM energy $\sqrt{s}$ deviates from the $Z^0$ pole even by only 1\%$m_Z$, which reflects the effects of ISR.

%%%%%%%%%%%%%%%%%%%%%%%
\begin{table}
\caption{The deviations of cross sections (units:~$\times 10^{-3}$ fb) for the top six channels for the production of double charmonia, double bottomonia, and double $B_c$ mesons when the CM energy $\sqrt{s}$ deviates from the $Z^0$ mass pole by 1\%$m_Z$ and 3\%$m_Z$. The top three channels are in bold face.}
\begin{tabular}{|c|c|c|c|c|c|}
\hline
~$\sqrt{s}$~&$97\%m_Z$&$99\%m_Z$&$m_Z$&$101\%m_Z$&$103\%m_Z$\\
\hline\hline
$\sigma{(J/\psi+J/\psi)}_{Z^0}$&~0.1279~&~0.4357~&0.6416&~0.4006~&~0.09582~\\
\hline
$\sigma{(\eta_c+h_c)}_{Z^0}$&~0.1139~&~0.4043~&0.6076&~0.3870~&~0.09626~\\
\hline
\bm{$\sigma{(J/\psi+h_c)}_{Z^0}$}&~\textbf{0.8560}~&~\textbf{3.026}~&\textbf{4.540}&~\textbf{2.886}~&~\textbf{0.7155}~\\
\hline
\bm{$\sigma{(\eta_c+\chi_{c0})}_{Z^0}$}&~\textbf{0.2567}~&~\textbf{0.9112}~&\textbf{1.370}&~\textbf{0.8723}~&~\textbf{0.2170}~\\
\hline
\bm{$\sigma{(\eta_c+\chi_{c2})}_{Z^0}$}&~\textbf{0.5339}~&~\textbf{1.892}~&\textbf{2.842}&~\textbf{1.809}~&~\textbf{0.4494}~\\
\hline
$\sigma{(J/\psi+\chi_{c2})}_{Z^0}$&~0.1116~&~0.3870~&0.5751&~0.3622~&~0.08821~\\
%%%%%%%%%%%%
\hline\hline
\bm{$\sigma{(\eta_b+\Upsilon)}_{Z^0}$}&\textbf{8.510}&\textbf{29.07}&\textbf{42.86}&~\textbf{26.79}~&~\textbf{6.421}~\\
\hline
\bm{$\sigma{(\Upsilon+\Upsilon)}_{Z^0}$}&\textbf{2.373}&\textbf{8.122}&\textbf{11.99}&~\textbf{7.498}~&~\textbf{1.800}~\\
\hline
$\sigma{(\Upsilon+h_b)}_{Z^0}$&1.182&4.124&6.146&~3.882~&~0.9501~\\
\hline
$\sigma{(\Upsilon+\chi_{b0})}_{Z^0}$&1.336&4.612&6.837&~4.296~&~1.041~\\
\hline
$\sigma{(\Upsilon+\chi_{b1})}_{Z^0}$&2.337&7.971&11.75&~7.337~&~1.757~\\
\hline
\bm{$\sigma{(\Upsilon+\chi_{b2})}_{Z^0}$}&\textbf{4.546}&\textbf{15.62}&\textbf{23.06}&~\textbf{14.45}~&~\textbf{3.482}~\\
%%%%%%%%%%%%
\hline\hline
\bm{$\sigma{(\eta_{bc}^+ + B_c^{*-})}_{Z^0}$}&~\textbf{121.9}~&~\textbf{425.6}~&\textbf{634.6}&~\textbf{401.0}~&~\textbf{98.27}~\\
\hline
\bm{$\sigma{(B_c^{*+} + B_c^{*-})}_{Z^0}$}&~\textbf{220.3}~&~\textbf{770.8}~&\textbf{1150}&~\textbf{727.5}~&~\textbf{178.6}~\\
\hline
$\sigma{(\eta_{bc}^+ +\chi_{bc0}^-)}_{Z^0}$&~2.851~&~10.15~&15.27&~9.739~&~2.429~\\
\hline
$\sigma{(B_c^{*+}+\chi_{bc0}^-)}_{Z^0}$&~81.63~&~298.0~&454.2&~293.3~&~75.03~\\
\hline
$\sigma{(B_c^{*+}+\chi_{bc1}^-)}_{Z^0}$&~51.16~&~180.3~&270.1&~171.5~&~42.40~\\
\hline
\bm{$\sigma{(B_c^{*+}+\chi_{bc2}^-)}_{Z^0}$}&~\textbf{20.57}~&~\textbf{745.6}~&\textbf{1133}&~\textbf{729.0}~&~\textbf{185.3}~\\
\hline\hline
\end{tabular}
\label{sqrtsdeviation}
\end{table}
%%%%%%%%%%%%%%%%%%%%%%%

The dependence of total cross sections on the renormalization scale $\mu$ in the running strong coupling $\alpha_s(\mu)$ is usually large in the leading order calculation.
With its interpretation as an estimate of the contribution from missing higher order, it might be indicative.
In the plots for total cross sections with running scale $\mu$ in Figs. \ref{ccds}-\ref{bcds}, we find that the cross sections always decrease monotonically as the scale increases with a more smooth trend at higher energy.

The differential angle distributions $d\sigma/dcos\theta$ for the double charmonia, double bottomonia and the double $B_c$ mesons are also displayed Figs. \ref{ccds}-\ref{bcds}, respectively.
The distribution $d\sigma/dcos\theta$ can be obtained easily with the help of the differential phase space in Eq. (\ref{dPhi-2}). Here, $\theta$ is the angle between the momentum $\vec{p_1}$ of the electron and the momentum $\vec{q_1}$ of the heavy quarkonium.
It is found that the distribution of the production of $\Upsilon+\chi_{b0}$ is quite flat, while other distributions have either a slight bulge or a slight hollow.
We can derive the $p_t$ distribution $d\sigma/dp_t$ with the help of angle distributions as follows
\begin{eqnarray}
\frac{d\sigma}{dp_t}=\left|\frac{dcos\theta}{dp_t}\right| \left(\frac{d\sigma}{dcos\theta}\right)%\nonumber\\
=\frac{p_t}{|\vec{q}_1| \sqrt{|\vec{q}_1|^2-p^2_t}} \left(\frac{d\sigma}{dcos\theta}\right),
\label{eq:dpt}
\end{eqnarray}
where ${|\vec{q}_1|}=\sqrt{\lambda[s, M^2_{1}, M^2_{2}]} /2\sqrt{s}$ is the magnitude of the 3-momentum of the heavy quarkonium.
As are shown in Figs. \ref{ccds}-\ref{bcds}, the differential cross sections versus $p_t$ of all channels are monotonically increasing.
This is understood because the $p_t$ distribution is proportional to $p_t/\sqrt{|\vec{q}_1|^2-p^2_t}$ and the distribution $d\sigma/dcos\theta$ changes relatively smoothly.

%%%%%%%%%%%%%%%%%%%%%%%%%%%%%%%%
\subsection{Uncertainties}
\label{uncertainty}
%%%%%%%%%%%%%%%%%%%%%%%%%%%%%%%%

%%%%%%%%%%%%%%%%%%%%%%%Uncertainties of total cross sections

\begin{table}
\caption{Uncertainties of the total cross sections (units:~$\times 10^{-5}~fb$) caused by constituent quark masses under B.T. model, and by four other potential models for $Z^0$ propagated channels of double charmonia at $\sqrt{s}=m_Z$. The percentages in brackets are the ratios of the maximum or minimum in other potential models to the estimates under the B.T. model.}
\begin{tabular}{|c|c|c|c|c|c|}
\hline
Potential Models & ~B.T.~&~J.~&~I.O.~&~C.K.~&~Cor.~\\
\hline\hline
$\sigma{(\eta_c+J/\psi)}_{Z^0}$&~$18.95^{+3.64}_{-4.70}$~&~3.927~&~1.001(5\%)~&~1.653 ~&~2.976~\\
\hline
$\sigma{(J/\psi+J/\psi)}_{Z^0}$&~$64.16^{+12.29}_{-15.87}$~&~13.30~&~3.390(5\%)~&~5.597 ~&~10.07~\\
\hline
$\sigma{(\eta_c+h_c)}_{Z^0}$&~$60.76^{+4.49}_{-7.35}$~&~14.78~&~2.299(4\%)~&~4.124~&~6.804~\\
\hline
$\sigma{(J/\psi+h_c)}_{Z^0}$&~$454.0^{+39.1}_{-59.2}$~&~110.4~&~17.18(4\%)~&~30.82~&~50.84~\\
\hline
$\sigma{(\eta_c+\chi_{c0})}_{Z^0}$&~$137.0^{+9.9}_{-16.4}$~&~33.32~&~5.183(4\%)~&~9.299~&~15.34~\\
\hline
$\sigma{(\eta_c+\chi_{c1})}_{Z^0}$&~$18.87^{+3.98}_{-4.27}$~&~4.589~&~0.714(4\%)~&~1.281~&~2.113~\\
\hline
$\sigma{(\eta_c+\chi_{c2})}_{Z^0}$&~$284.2^{+22.2}_{-35.3}$~&~69.11~&~10.75(4\%)~&~19.29~&~31.83~\\
\hline
$\sigma{(J/\psi+\chi_{c0})}_{Z^0}$&~$44.03^{+4.60}_{-6.38}$~&~10.71~&~1.666(4\%)~&~2.989~&~4.931~\\
\hline
$\sigma{(J/\psi+\chi_{c1})}_{Z^0}$&~$17.62^{+3.57}_{-3.88}$~&~4.285~&~0.667(4\%)~&~1.196~&~1.973~\\
\hline
$\sigma{(J/\psi+\chi_{c2})}_{Z^0}$&~$57.51^{+8.25}_{-10.02}$~&~13.99~&~2.176(4\%)~&~3.903~&~6.440~\\
\hline
\end{tabular}
\label{tabrpd}
\end{table}
%%%%%%%%%%%%%%%%%%%%%%%

%%%%%%%%%%%%%%%%%%%%%%%
\begin{table}
\caption{Uncertainties of the total cross sections (units:~$\times 10^{-4}~fb$) caused by constituent quark masses under B.T. model, and by four other potential models for $Z^0$ propagated channels of double bottomonia at $\sqrt{s}=m_Z$. The percentages in brackets are the ratios of the maximum or minimum in other potential models to the estimates under the B.T. model.}
\begin{tabular}{|c|c|c|c|c|c|}
\hline
Potential Models &~B.T.~&~J.~&~I.O.~&~C.K.~&~Cor.~\\
\hline\hline
$\sigma{(\eta_b+\Upsilon)}_{Z^0}$&~$428.6^{+67.9}_{-60.9}$~&~83.47~&~164.3~&~46.30(11\%)~&~137.8~\\
\hline
$\sigma{(\Upsilon+\Upsilon)}_{Z^0}$&~$119.9^{+18.4}_{-16.6}$~&~23.35~&~45.97~&~12.95(11\%)~&~38.55~\\
\hline
$\sigma{(\eta_b+h_{b})}_{Z^0}$&~$53.27^{+1.46}_{-1.51}$~&~6.580~&~6.542~&~3.311(6\%)~&~6.263~\\
\hline
$\sigma{(\Upsilon+h_{b})}_{Z^0}$&~$61.46^{+4.22}_{-4.05}$~&~7.591~&~7.547~&~3.820(6\%)~&~7.226~\\
\hline
$\sigma{(\eta_b+\chi_{b0})}_{Z^0}$&~$9.215^{+0.141}_{-0.160}$~&~1.138~&~1.132~&~0.573(6\%)~&~1.083~\\
\hline
$\sigma{(\eta_b+\chi_{b1})}_{Z^0}$&~$14.31^{+1.60}_{-1.60}$~&~1.767~&~1.757~&~0.890(6\%)~&~1.682~\\
\hline
$\sigma{(\eta_b+\chi_{b2})}_{Z^0}$&~$28.15^{+1.38}_{-1.35}$~&~3.477~&~3.457~&~1.750(6\%)~&~3.310~\\
\hline
$\sigma{(\Upsilon+\chi_{b0})}_{Z^0}$&~$68.37^{+6.17}_{-5.83}$~&~8.445~&~8.396~&~4.250(6\%)~&~8.038~\\
\hline
$\sigma{(\Upsilon+\chi_{b1})}_{Z^0}$&~$117.5^{+13.6}_{-15.0}$~&~14.51~&~14.43~&~7.304(6\%)~&~13.81~\\
\hline
$\sigma{(\Upsilon+\chi_{b2})}_{Z^0}$&~$230.6^{+23.6}_{-22.2}$~&~28.48~&~28.32~&~14.33(6\%)~&~27.11~\\
\hline
\end{tabular}
\label{tabrpe}
\end{table}
%%%%%%%%%%%%%%%%%%%%%%%

%%%%%%%%%%%%%%%%%%%%%%%
\begin{table}
\caption{Uncertainties of the total cross sections (units:~$\times 10^{-3}~fb$) caused by constituent quark masses under B.T. model, and by four other potential models for $Z^0$ propagated channels of the double $B_c$ mesons at $\sqrt{s}=m_Z$. The percentages in brackets are the ratios of the maximum or minimum in other potential models to the estimates under the B.T. model.}
\begin{tabular}{|c|c|c|c|c|c|}
\hline
Potential Models &~B.T.~&~J.~&~I.O.~&~C.K.~&~Cor.~ \\
\hline\hline
$\sigma{(\eta_{bc}^++B_c^{*-})}_{Z^0}$&~$634.6^{+47.8}_{-45.8}$~&~175.1~&~1653(260\%)~&~72.88(11\%)~&~136.2~\\
\hline
$\sigma{(B_c^{*+}+B_c^{*-})}_{Z^0}$&~$1150^{+87}_{-84}$~&~317.2~&~2996(261\%)~&~132.1(11\%)~&~246.9~\\
\hline
$\sigma{(\eta_{bc}^++h_{bc}^-)}_{Z^0}$&~$4.810^{+0.399}_{-0.406}$~&~2.014~&~8.588(179\%)~&~0.585(12\%)~&~0.942~\\
\hline
$\sigma{(B_c^{*+}+h_{bc}^-)}_{Z^0}$&~$11.77^{+1.10}_{-1.05}$~&~4.929~&~21.01(179\%)~&~1.432(12\%)~&~2.306~\\
\hline
$\sigma{(\eta_{bc}^++\chi_{bc0}^-)}_{Z^0}$&~$15.27^{+1.10}_{-1.12}$~&~6.394~&~27.26(179\%)~&~1.858(12\%)~&~2.991~\\
\hline
$\sigma{(\eta_{bc}^++\chi_{bc1}^-)}_{Z^0}$&~$6.162^{+0.701}_{-0.448}$~&~2.580~&~11.00(179\%)~&~0.750(12\%)~&~1.207~\\
\hline
$\sigma{(\eta_{bc}^++\chi_{bc2}^-)}_{Z^0}$&~$3.093^{+0.253}_{-0.236}$~&~1.295~&~5.522(179\%)~&~0.376(12\%)~&~0.606~\\
\hline
$\sigma{(B_c^{*+}+\chi_{bc0}^-)}_{Z^0}$&~$454.2^{-15.7}_{+30.8}$~&~190.2~&~811.0(179\%)~&~55.27(12\%)~&~88.98~\\
\hline
$\sigma{(B_c^{*+}+\chi_{bc1}^-)}_{Z^0}$&~$270.1^{+16.2}_{-15.8}$~&~113.1~&~482.3(179\%)~&~32.87(12\%)~&~52.91~\\
\hline
$\sigma{(B_c^{*+}+\chi_{bc2}^-)}_{Z^0}$&~$1133^{-16}_{+22}$~&~474.4~&~2023(179\%)~&~137.9(12\%)~&~222.0~\\
\hline
\end{tabular}
\label{tabrpf}
\end{table}
%%%%%%%%%%%%%%%%%%%%%%%

To obtain reliable estimates at leading order calculation, we discuss the main uncertainty sources of the cross sections of the double quarkonium or $B_c$ production in this subsection.
For the input parameters, the Fermi constant $G_F$, the fine structure constant $\alpha$, the Weinberg angle $\theta_W$, and the mass and width of the $Z^0$ boson are either relatively precise or an overall factor.
So we shall explore the uncertainties caused by masses of constituent heavy quarks, the radial wave functions or their first derivatives at the origin\footnote{The radial wave functions or their first derivatives at the origin can be related to the inclusive decay of heavy quarkonia or $B_c$ mesons, and parts of them can be extracted from experimental measurements. However, we can not obtain all the values from experiments. Thus we discuss the uncertainties from different theoretical potential models.}.
We only discuss the uncertainties of the $Z^0$ propagated channels since the contributions of virtual photon channels are smaller by three orders of magnitude at least at super $Z$ factory.

The uncertainties of cross sections caused by both varying quark masses and the radial wave functions or their first derivatives at the origin at $\sqrt{s}=m_Z$ are presented in Tables~\ref{tabrpd}-\ref{tabrpf} for the double charmonia, double bottomonia, and double $B_c$ mesons, respectively.
The uncertainties caused by varying masses are displayed in the second columns which are calculated in the BT-potential model~\cite{lx}, and the mass deviations are 0.1 GeV for $m_c$ and 0.2 GeV for $m_b$ (see explicit values in Table \ref{M&R}).
Here, the affects of the varying masses on the radial wave functions or their first derivatives at the origin are also taken into consideration, which makes them not being the overall factors any more.
The values from the third to the sixth columns in Tables~\ref{tabrpd}-\ref{tabrpf} are the estimates with the radial wave functions or their first derivatives at the origin calculated in other four potential models.
The four models are the QCD-motivated potential with one-loop correction given by John L. Richardson (J. potential) \cite{Richardson:1978bt},
the QCD-motivated potential with two-loop correction given by K. Igi and S. Ono (I.O. potential) \cite{IO,Ikhdair:2003ry},
the QCD-motivated potential with two-loop correction given by Yu-Qi Chen and Yu-Ping Kuang (C.K. potential) \cite{Chen:1992fq,Ikhdair:2003ry},
and the QCD-motivated Coulomb-plus-linear potential (Cor. potential) \cite{Ikhdair:2003ry,Eichten:1978tg,Eichten:1979ms,Eichten:1980mw,Eichten:1995ch}.
The formula and latest values of those wave functions or their first derivatives at the origin have been discussed in our earlier work \cite{lx}.
It is shown that the cross sections change dramatically when we choose different potential models, which bring the largest uncertainties.
For the production of the double charmonia, we always obtain the maximum under the B.T. potential model and obtain the minimum under the I.O. potential model.
For the production of the double bottomonia, we always obtain the maximum under the B.T. potential model and obtain the minimum under the C.K. potential model.
For the production of the double $B_c$ mesons, we always obtain the maximum under the I.O. potential model and obtain the minimum under the C.K. potential model.
In the Tables~\ref{tabrpd}-\ref{tabrpf}, percentages in brackets are the ratios of the maximum or minimum relative to the estimates under the B.T. model.

\section{Summery}
%%%%%%%%%%%%%%%%
%%%%%%%%%%%%%%%%%%%%%%%%%%%%%%%%
In the present work, we study the production of the double heavy quarkonia or $B_c$ mesons in $e^-(p_1) e^+(p_2) \to \gamma^*/Z^0 \to H_1[n](q_1) + H_2[n^\prime](q_2)$ within the CSM framework, where $H_{1(2)}$ represent the heavy quarkonia or $B_c$ mesons and $[n^{(\prime)}]$ represent the color-singlet Fock states $[^1S_0], ~[^3S_1], ~[^1P_1]$, and $[^3P_J]$ ($J=0,1,2$).
For a sound estimation at the future super $Z$ factory, total cross sections $\sigma$ as functions of CM energy $\sqrt{s}$, $\sigma$ as functions of the renormalization scale $\mu$, the angle distributions $d\sigma/dcos\theta$,  and the $p_T$ distributions $d\sigma/dp_{t}$ are studied, which are shown in Figs. \ref{ccds}-\ref{bcds} for the double charmonia, double bottomonia, and double $B_c$ mesons, respectively.
We further discuss the uncertainties of the cross sections caused by the varying masses of constituent quarks, and the non-perturbative matrix elements under different potential models.

In Tables \ref{tabrpa}, it is found that the production rates of double $B_c$ meson channels are roughly one order of magnitude greater than those of double bottomonium channels with the same Fock states, and are roughly two orders of magnitude greater than those of the double charmonium channels with the same Fock states.
For CEPC running in $Z$ factory mode, we obtain $\mathcal{O}(1)$ events of $J/\psi+h_c$ signal because of the big suppression factors of the fully reconstructed of both heavy charmonia.
For the $Z$ factory mode at FCC-ee, we might catch over a dozen of such rare events, but it is pessimistic for the observation in experiments after taking the reconstruction efficiency into account.
However, since the color-singlet model predictions often turn out to undershoot data by one or even more orders of magnitude in many quarkonium production processes,
therefore a detection of the considered processes at the FCC-ee and also the CEPC is still far from excluded.
As shown in Table \ref{sqrtsdeviation}, the ISR may bring about 30\%$\sim$40\%  suppresions when the CM energy $\sqrt{s}$ deviates from the $Z^0$ pole even by only 1\%$m_Z$.
For the uncertainty analyses in Tables \ref{tabrpd}-\ref{tabrpf}, we find that the masses of constituent quarks can bring up to about 20\% corrections to the cross sections.
And the total cross sections can increase by about $2\sim3$ times or decrease by an order of magnitude when adopting different potential models, which would be the major source of uncertainty.

\hspace{2cm}

{\bf Acknowledgements}:
This work was supported in part by
the National Natural Science Foundation of China under Grant No. 11905112,
and the Natural Science Foundation of Shandong Province under Grant No. ZR2019QA012.


\begin{thebibliography}{99}

%%%%%%%%%%%%%%%%%%%%%%%%%%%%%
%INTRODUCTION
%%%%%%%%%%%%%%%%%%%%%%%%%%%%%

%\cite{CEPCStudyGroup:2018ghi}
\bibitem{CEPCStudyGroup:2018ghi}
J.~B.~Guimar\~aes da Costa \textit{et al.} [CEPC Study Group],
``CEPC Conceptual Design Report: Volume 2 - Physics \& Detector,''
[arXiv:1811.10545 [hep-ex]].
%414 citations counted in INSPIRE as of 28 May 2022

%\cite{Agapov:2022bhm}
\bibitem{Agapov:2022bhm}
I.~Agapov, M.~Benedikt, A.~Blondel, M.~Boscolo, O.~Brunner, M.~C.~Llatas, T.~Charles, D.~Denisov, W.~Fischer and E.~Gianfelice-Wendt, \textit{et al.}
``Future Circular Lepton Collider FCC-ee: Overview and Status,''
[arXiv:2203.08310 [physics.acc-ph]].
%9 citations counted in INSPIRE as of 12 Sep 2022

%\cite{ECFADESYLCPhysicsWorkingGroup:2001igx}
\bibitem{ECFADESYLCPhysicsWorkingGroup:2001igx}
J.~A.~Aguilar-Saavedra \textit{et al.} [ECFA/DESY LC Physics Working Group],
``TESLA: The Superconducting electron positron linear collider with an integrated x-ray laser laboratory. Technical design report. Part 3. Physics at an e+ e- linear collider,''
[arXiv:hep-ph/0106315 [hep-ph]].
%1090 citations counted in INSPIRE as of 14 Sep 2021

%Z factory
\bibitem{jz} J.~P.~Ma and Z.~X.~Zhang,
``Preface,''
Sci. China Phys. Mech. Astron. 53, issue 11, 1947-1948 (2010).
%https://doi.org/10.1007/s11433-010-4144-5

%heavy quarkonium
%
%\cite{Brambilla:2010cs}
\bibitem{Brambilla:2010cs}
N.~Brambilla, S.~Eidelman, B.~K.~Heltsley, R.~Vogt, G.~T.~Bodwin, E.~Eichten, A.~D.~Frawley, A.~B.~Meyer, R.~E.~Mitchell and V.~Papadimitriou, \textit{et al.}
``Heavy Quarkonium: Progress, Puzzles, and Opportunities,''
Eur. Phys. J. C \textbf{71}, 1534 (2011),
%doi:10.1140/epjc/s10052-010-1534-9
[arXiv:1010.5827 [hep-ph]].
%1624 citations counted in INSPIRE as of 15 Sep 2021

%\cite{Andronic:2015wma}
\bibitem{Andronic:2015wma}
A.~Andronic, F.~Arleo, R.~Arnaldi, A.~Beraudo, E.~Bruna, D.~Caffarri, Z.~C.~del Valle, J.~G.~Contreras, T.~Dahms and A.~Dainese, \textit{et al.}
``Heavy-flavour and quarkonium production in the LHC era: from proton\textendash{}proton to heavy-ion collisions,''
Eur. Phys. J. C \textbf{76}, no.3, 107 (2016),
%doi:10.1140/epjc/s10052-015-3819-5
[arXiv:1506.03981 [nucl-ex]].
%517 citations counted in INSPIRE as of 02 Jun 2022

%\cite{Chung:2018lyq}
\bibitem{Chung:2018lyq}
H.~S.~Chung,
``Review of quarkonium production: status and prospects,''
PoS \textbf{Confinement2018}, 007 (2018),
%doi:10.22323/1.336.0007
[arXiv:1811.12098 [hep-ph]].
%6 citations counted in INSPIRE as of 15 Sep 2021

%\cite{Chen:2021tmf}
\bibitem{Chen:2021tmf}
A.~P.~Chen, Y.~Q.~Ma and H.~Zhang,
``A Short Theoretical Review of Charmonium Production,''
Adv. High Energy Phys. \textbf{2022}, 7475923 (2022),
%doi:10.1155/2022/7475923
[arXiv:2109.04028 [hep-ph]].
%7 citations counted in INSPIRE as of 07 Aug 2022

%\cite{Chapon:2020heu}
\bibitem{Chapon:2020heu}
E.~Chapon, D.~d'Enterria, B.~Ducloue, M.~G.~Echevarria, P.~B.~Gossiaux, V.~Kartvelishvili, T.~Kasemets, J.~P.~Lansberg, R.~McNulty and D.~D.~Price, \textit{et al.}
``Prospects for quarkonium studies at the high-luminosity LHC,''
Prog. Part. Nucl. Phys. \textbf{122}, 103906 (2022),
%doi:10.1016/j.ppnp.2021.103906
[arXiv:2012.14161 [hep-ph]].
%47 citations counted in INSPIRE as of 02 Jun 2022


%nrqcd
\bibitem{nrqcd1}
  G.~T.~Bodwin, E.~Braaten and G.~P.~Lepage,
  ``Rigorous QCD analysis of inclusive annihilation and production of heavy quarkonium,''
  Phys.\ Rev.\ D {\bf 51}, 1125 (1995),
  Erratum: [Phys.\ Rev.\ D {\bf 55}, 5853 (1997)].
\bibitem{nrqcd2}
 A.~Petrelli, M.~Cacciari, M.~Greco, F.~Maltoni and M.~L.~Mangano,
  ``NLO production and decay of quarkonium,''
  Nucl.\ Phys.\ B {\bf 514}, 245 (1998).

%csm
%\cite{Ellis:1976fj}
\bibitem{Ellis:1976fj}
S.~D.~Ellis, M.~B.~Einhorn and C.~Quigg,
``Comment on Hadronic Production of Psions,''
Phys. Rev. Lett. \textbf{36}, 1263 (1976).
%doi:10.1103/PhysRevLett.36.1263
%166 citations counted in INSPIRE as of 10 Dec 2022

%\cite{Carlson:1976cd}
\bibitem{Carlson:1976cd}
C.~E.~Carlson and R.~Suaya,
``Hadronic Production of psi/J Mesons,''
Phys. Rev. D \textbf{14}, 3115 (1976).
%doi:10.1103/PhysRevD.14.3115
%141 citations counted in INSPIRE as of 10 Dec 2022

%\cite{Chang:1979nn}
\bibitem{Chang:1979nn}
C.~H.~Chang,
``Hadronic Production of $J/\psi$ Associated With a Gluon,''
Nucl. Phys. B \textbf{172}, 425-434 (1980).
%doi:10.1016/0550-3213(80)90175-3
%344 citations counted in INSPIRE as of 10 Dec 2022

%\cite{wgs,pot2,Richardson:1978bt,IO,Ikhdair:2003ry,Chen:1992fq,Eichten:1978tg,Eichten:1979ms,Eichten:1980mw,Eichten:1995ch}
%BT-potential
\bibitem{wgs}
W.~Buchmuller, G.~Grunberg and S.~H.~H.~Tye,
``The Regge Slope and the Lambda Parameter in QCD: An Empirical Approach via Quarkonia,''
Phys. Rev. Lett. \textbf{45}, 103 (1980),
[erratum: Phys. Rev. Lett. \textbf{45}, 587 (1980)].
%
\bibitem{pot2}
%\bibitem{Buchmuller:1980su}
W.~Buchmuller and S.~H.~H.~Tye,
``Quarkonia and Quantum Chromodynamics,''
Phys. Rev. D \textbf{24}, 132 (1981).
%doi:10.1103/PhysRevD.24.132

%J potential
%\cite{Richardson:1978bt}
\bibitem{Richardson:1978bt}
J.~L.~Richardson,
``The Heavy Quark Potential and the Upsilon, J/psi Systems,''
Phys. Lett. B \textbf{82}, 272-274 (1979).

%IO potential, CK potential & Cor potential
%\cite{IO}
\bibitem{IO}
K.~Igi and S.~Ono,
``Heavy Quarkonium Systems and the {QCD} Scale Parameter $\Lambda$ Ms,''
Phys. Rev. D \textbf{33}, 3349 (1986)
%doi:10.1103/PhysRevD.33.3349
%82 citations counted in INSPIRE as of 02 Jun 2022

%\cite{Ikhdair:2003ry}
\bibitem{Ikhdair:2003ry}
S.~M.~Ikhdair and R.~Sever,
``Spectroscopy of $B_c$ meson in a semirelativistic quark model using the shifted large $N$ expansion method,''
Int. J. Mod. Phys. A \textbf{19}, 1771-1792 (2004).

%CK potential
%\cite{Chen:1992fq}
\bibitem{Chen:1992fq}
Y.~Q.~Chen and Y.~P.~Kuang,
``Improved QCD motivated heavy quark potentials with explicit Lambda(ms) dependence,''
Phys. Rev. D \textbf{46}, 1165 (1992),
[erratum: Phys. Rev. D \textbf{47}, 350 (1993)].

%Cor potential
\bibitem{Eichten:1978tg}
  E.~Eichten, K.~Gottfried, T.~Kinoshita, K.~D.~Lane and T.~M.~Yan,
  ``Charmonium: The Model,''
  Phys.\ Rev.\ D {\bf 17}, 3090 (1978),
  Erratum: [Phys.\ Rev.\ D {\bf 21}, 313 (1980)].

\bibitem{Eichten:1979ms}
  E.~Eichten, K.~Gottfried, T.~Kinoshita, K.~D.~Lane and T.~M.~Yan,
  ``Charmonium: Comparison with Experiment,''
  Phys.\ Rev.\ D {\bf 21}, 203 (1980).

%\cite{Eichten:1980mw}
\bibitem{Eichten:1980mw}
E.~Eichten and F.~Feinberg,
``Spin Dependent Forces in QCD,''
Phys. Rev. D \textbf{23}, 2724 (1981).
%doi:10.1103/PhysRevD.23.2724
%818 citations counted in INSPIRE as of 22 Sep 2021

%\cite{Eichten:1995ch}
\bibitem{Eichten:1995ch}
E.~J.~Eichten and C.~Quigg,
``Quarkonium wave functions at the origin,''
Phys. Rev. D \textbf{52}, 1726-1728 (1995).



%pNRQCD
%\cite{Brambilla:1999xf}
\bibitem{Brambilla:1999xf}
N.~Brambilla, A.~Pineda, J.~Soto and A.~Vairo,
``Potential NRQCD: An Effective theory for heavy quarkonium,''
Nucl. Phys. B \textbf{566}, 275 (2000)
%doi:10.1016/S0550-3213(99)00693-8
[arXiv:hep-ph/9907240 [hep-ph]].
%741 citations counted in INSPIRE as of 11 Dec 2022

%lattice
%\cite{Bodwin:1996tg}
\bibitem{Bodwin:1996tg}
G.~T.~Bodwin, D.~K.~Sinclair and S.~Kim,
``Quarkonium decay matrix elements from quenched lattice QCD,''
Phys. Rev. Lett. \textbf{77}, 2376-2379 (1996)
%doi:10.1103/PhysRevLett.77.2376
[arXiv:hep-lat/9605023 [hep-lat]].
%94 citations counted in INSPIRE as of 11 Dec 2022

%trace technology
\bibitem{cjx}
C.~H.~Chang, J.~X.~Wang and X.~G.~Wu,
  ``Production of $B_c$ or $\bar{B}_c$ meson and its excited states in $\bar{t}$ quark or $t$ quark decays,''
  Phys.\ Rev.\ D {\bf 77}, 014022 (2008).
\bibitem{lxz}
L.~C.~Deng, X.~G.~Wu, Z.~Yang, Z.~Y.~Fang and Q.~L.~Liao,
  ``$Z_0$ Boson Decays to $B^{(*)}_c$ Meson and Its Uncertainties,''
  Eur.\ Phys.\ J.\ C {\bf 70}, 113 (2010).
%\cite{Yang:2011ps}
\bibitem{Yang:2011ps}
Z.~Yang, X.~G.~Wu, G.~Chen, Q.~L.~Liao and J.~W.~Zhang,
``$B_c$ Meson Production around the $Z^0$ Peak at a High Luminosity $e^+ e^-$ Collider,''
Phys. Rev. D \textbf{85}, 094015 (2012),
%doi:10.1103/PhysRevD.85.094015
[arXiv:1112.5169 [hep-ph]].
%15 citations counted in INSPIRE as of 22 Sep 2021
\bibitem{wbc1} Q.~L.~Liao, X.~G.~Wu, J.~Jiang, Z.~Yang and Z.~Y.~Fang,
  ``Heavy Quarkonium Production at LHC through $W$ Boson Decays,''
  Phys.\ Rev.\ D {\bf 85}, 014032 (2012).
 %\cite{Liao:2012rh}
\bibitem{Liao:2012rh}
Q.~L.~Liao, X.~G.~Wu, J.~Jiang, Z.~Yang, Z.~Y.~Fang and J.~W.~Zhang,
``Excited Heavy Quarkonium Production at the LHC through $W$-Boson Decays,''
Phys. Rev. D \textbf{86}, 014031 (2012),
%doi:10.1103/PhysRevD.86.014031
[arXiv:1204.2594 [hep-ph]].
%11 citations counted in INSPIRE as of 02 Jun 2022
%\cite{Liao:2021ifc}
\bibitem{Liao:2021ifc}
Q.~L.~Liao, J.~Jiang, P.~C.~Lu and G.~Chen,
``Production of excited heavy quarkonia in $e^+e^-\to \gamma^*/Z^0 \to |(Q \bar{Q})[n]\rangle+ \gamma$ at super Z factory,
Phys. Rev. D \textbf{105} (2022) no.1, 016026
%doi:10.1103/PhysRevD.105.016026
[arXiv:2112.03522 [hep-ph]].
%0 citations counted in INSPIRE as of 17 Apr 2022


% double quarkonium production in pp collision
%%%
%\cite{Schafer:2019ynn}
\bibitem{Schafer:2019ynn}
W.~Sch\"afer,
``Production of quarkonium pairs in high-energy proton-proton collisions,''
EPJ Web Conf. \textbf{199}, 01021 (2019).
%doi:10.1051/epjconf/201919901021
%4 citations counted in INSPIRE as of 02 Jun 2022
%\cite{Scarpa:2019fol}
\bibitem{Scarpa:2019fol}
F.~Scarpa, D.~Boer, M.~G.~Echevarria, J.~P.~Lansberg, C.~Pisano and M.~Schlegel,
``Studies of gluon TMDs and their evolution using quarkonium-pair production at the LHC,''
Eur. Phys. J. C \textbf{80}, no.2, 87 (2020),
%doi:10.1140/epjc/s10052-020-7619-1
[arXiv:1909.05769 [hep-ph]].
%31 citations counted in INSPIRE as of 02 Jun 2022
%\cite{He:2019qqr}
\bibitem{He:2019qqr}
Z.~G.~He, B.~A.~Kniehl, M.~A.~Nefedov and V.~A.~Saleev,
``Double Prompt $J/\psi$ Hadroproduction in the Parton Reggeization Approach with High-Energy Resummation,''
Phys. Rev. Lett. \textbf{123}, no.16, 162002 (2019),
%doi:10.1103/PhysRevLett.123.162002
[arXiv:1906.08979 [hep-ph]].
%20 citations counted in INSPIRE as of 02 Jun 2022
%\cite{Qiao:2002rh}
\bibitem{Qiao:2002rh}
C.~F.~Qiao,
``$J/\psi$ pair production at the Tevatron,''
Phys. Rev. D \textbf{66}, 057504 (2002)
%doi:10.1103/PhysRevD.66.057504
[arXiv:hep-ph/0206093 [hep-ph]].
%33 citations counted in INSPIRE as of 02 Jun 2022
%\cite{Qiao:2009kg}
\bibitem{Qiao:2009kg}
C.~F.~Qiao, L.~P.~Sun and P.~Sun,
``Testing Charmonium Production Mechamism via Polarized J/psi Pair Production at the LHC,''
J. Phys. G \textbf{37}, 075019 (2010),
%doi:10.1088/0954-3899/37/7/075019
[arXiv:0903.0954 [hep-ph]].
%60 citations counted in INSPIRE as of 02 Jun 2022
%\cite{Lansberg:2013qka}
\bibitem{Lansberg:2013qka}
J.~P.~Lansberg and H.~S.~Shao,
``Production of $J/\psi + \eta_{c}$ versus $J/\psi + J/\psi$ at the LHC: Importance of Real $\alpha^{5}_{s}$ Corrections,''
Phys. Rev. Lett. \textbf{111}, 122001 (2013),
%doi:10.1103/PhysRevLett.111.122001
[arXiv:1308.0474 [hep-ph]].
%94 citations counted in INSPIRE as of 02 Jun 2022
%\cite{Lansberg:2020rft}
\bibitem{Lansberg:2020rft}
J.~P.~Lansberg, H.~S.~Shao, N.~Yamanaka, Y.~J.~Zhang and C.~No\^us,
``Complete NLO QCD study of single- and double-quarkonium hadroproduction in the colour-evaporation model at the Tevatron and the LHC,''
Phys. Lett. B \textbf{807}, 135559 (2020),
%doi:10.1016/j.physletb.2020.135559
[arXiv:2004.14345 [hep-ph]].
%17 citations counted in INSPIRE as of 02 Jun 2022
%\cite{Lu:2021gxf}
\bibitem{Lu:2021gxf}
C.~Y.~Lu, D.~D.~Shen, P.~Sun and R.~Zhu,
``Soft Gluon Resummation in Double Heavy Quarkonium Production at LHC,''
[arXiv:2104.09941 [hep-ph]].
%0 citations counted in INSPIRE as of 02 Jun 2022

%photoproduction of double $J/\psi$ \cite{Xue-An:2018wat}
%\cite{Xue-An:2018wat}
\bibitem{Xue-An:2018wat}
P.~Xue-An, L.~Gang, S.~Mao, Z.~Yu, S.~Hao and G.~Jian-You,
``Photoproduction of the double $J/\psi$ ($\Upsilon$) at the LHC with forward proton taggin,''
Phys. Rev. D \textbf{99}, 014029 (2019),
%doi:10.1103/PhysRevD.99.014029
[arXiv:1812.08599 [hep-ph]].
%5 citations counted in INSPIRE as of 02 Jun 2022

%double heavy quarkonium produced in diffractive interactions
%\cite{BrennerMariotto:2018eef}
\bibitem{BrennerMariotto:2018eef}
C.~Brenner Mariotto, V.~P.~Gon\c{c}alves and R.~Palota da Silva,
``Double heavy quarkonium production in diffractive processes at the Run 2 LHC energy,''
Phys. Rev. D \textbf{98}, no.1, 014028 (2018),
%doi:10.1103/PhysRevD.98.014028
[arXiv:1806.00440 [hep-ph]].
%5 citations counted in INSPIRE as of 02 Jun 2022

% photon-photon interaction
%\cite{Chen:2020dtu}
\bibitem{Chen:2020dtu}
Z.~Q.~Chen, H.~Yang and C.~F.~Qiao,
``NLO QCD corrections to $B_c$-pair production in photon-photon collision,''
Phys. Rev. D \textbf{102}, no.1, 016011 (2020),
%doi:10.1103/PhysRevD.102.016011
[arXiv:2005.07317 [hep-ph]].
%4 citations counted in INSPIRE as of 06 Jun 2022
%\cite{Yang:2020xkl}
\bibitem{Yang:2020xkl}
H.~Yang, Z.~Q.~Chen and C.~F.~Qiao,
``NLO QCD corrections to exclusive quarkonium-pair production in photon\textendash{}photon collision,''
Eur. Phys. J. C \textbf{80}, no.9, 806 (2020),
%doi:10.1140/epjc/s10052-020-8390-z
%2 citations counted in INSPIRE as of 06 Jun 2022

% hadronic production of double $B_c$ mesons
%\cite{Li:2009ug}
\bibitem{Li:2009ug}
R.~Li, Y.~J.~Zhang and K.~T.~Chao,
``Pair Production of Heavy Quarkonium and B(c)(*) Mesons at Hadron Colliders,''
Phys. Rev. D \textbf{80}, 014020 (2009),
%doi:10.1103/PhysRevD.80.014020
[arXiv:0903.2250 [hep-ph]].
%47 citations counted in INSPIRE as of 02 Jun 2022


%Jpsi+etac issue
%
%\cite{Braaten:2002fi}
\bibitem{Braaten:2002fi}
E.~Braaten and J.~Lee,
``Exclusive Double Charmonium Production from $e^+ e^-$ Annihilation into a Virtual Photon,''
Phys. Rev. D \textbf{67}, 054007 (2003)
[erratum: Phys. Rev. D \textbf{72}, 099901 (2005)]
%doi:10.1103/PhysRevD.72.099901
[arXiv:hep-ph/0211085 [hep-ph]].
%272 citations counted in INSPIRE as of 02 Jun 2022
%\cite{Liu:2002wq}
\bibitem{Liu:2002wq}
K.~Y.~Liu, Z.~G.~He and K.~T.~Chao,
``Problems of double charm production in e+ e- annihilation at s**(1/2) = 10.6-GeV,''
Phys. Lett. B \textbf{557}, 45-54 (2003),
%doi:10.1016/S0370-2693(03)00176-X
[arXiv:hep-ph/0211181 [hep-ph]].
%188 citations counted in INSPIRE as of 02 Jun 2022
%\cite{Hagiwara:2003cw}
\bibitem{Hagiwara:2003cw}
K.~Hagiwara, E.~Kou and C.~F.~Qiao,
``Exclusive $J/\psi$ productions at $e^{+} e^{-}$ colliders,''
Phys. Lett. B \textbf{570}, 39-45 (2003),
%doi:10.1016/j.physletb.2003.07.006
[arXiv:hep-ph/0305102 [hep-ph]].
%106 citations counted in INSPIRE as of 02 Jun 2022

%\cite{Belle:2002tfa}
\bibitem{Belle:2002tfa}
K.~Abe \textit{et al.} [Belle],
``Observation of double c anti-c production in e+ e- annihilation at s**(1/2) approximately 10.6-GeV,''
Phys. Rev. Lett. \textbf{89}, 142001 (2002),
%doi:10.1103/PhysRevLett.89.142001
[arXiv:hep-ex/0205104 [hep-ex]].
%374 citations counted in INSPIRE as of 02 Jun 2022
%\cite{BaBar:2005nic}
\bibitem{BaBar:2005nic}
B.~Aubert \textit{et al.} [BaBar],
``Measurement of double charmonium production in $e^+e^-$ annihilations at $\sqrt{s}=10.6$ GeV,''
Phys. Rev. D \textbf{72}, 031101 (2005),
%doi:10.1103/PhysRevD.72.031101
[arXiv:hep-ex/0506062 [hep-ex]].
%227 citations counted in INSPIRE as of 02 Jun 2022

%\cite{Zhang:2005cha}
\bibitem{Zhang:2005cha}
Y.~J.~Zhang, Y.~j.~Gao and K.~T.~Chao,
``Next-to-leading order QCD correction to e+ e- ---\ensuremath{>} J / psi + eta(c) at s**(1/2) = 10.6-GeV,''
Phys. Rev. Lett. \textbf{96}, 092001 (2006),
%doi:10.1103/PhysRevLett.96.092001
[arXiv:hep-ph/0506076 [hep-ph]].
%188 citations counted in INSPIRE as of 02 Jun 2022
%\cite{Gong:2007db}
\bibitem{Gong:2007db}
B.~Gong and J.~X.~Wang,
``QCD corrections to $J/\psi$ plus $\eta_c$ production in $e^{+} e^{-}$ annihilation at $S^{(1/2)}$ = 10.6-GeV,''
Phys. Rev. D \textbf{77}, 054028 (2008),
%doi:10.1103/PhysRevD.77.054028
[arXiv:0712.4220 [hep-ph]].
%107 citations counted in INSPIRE as of 02 Jun 2022

%\cite{He:2007te}
\bibitem{He:2007te}
Z.~G.~He, Y.~Fan and K.~T.~Chao,
``Relativistic corrections to J/psi exclusive and inclusive double charm production at B factories,''
Phys. Rev. D \textbf{75}, 074011 (2007),
%doi:10.1103/PhysRevD.75.074011
[arXiv:hep-ph/0702239 [hep-ph]].
%123 citations counted in INSPIRE as of 02 Jun 2022
%\cite{Bodwin:2007ga}
\bibitem{Bodwin:2007ga}
G.~T.~Bodwin, J.~Lee and C.~Yu,
``Resummation of Relativistic Corrections to e+ e- ---\ensuremath{>} J/psi + eta(c),''
Phys. Rev. D \textbf{77}, 094018 (2008),
%doi:10.1103/PhysRevD.77.094018
[arXiv:0710.0995 [hep-ph]].
%114 citations counted in INSPIRE as of 02 Jun 2022

%\cite{Feng:2019zmt}
\bibitem{Feng:2019zmt}
F.~Feng, Y.~Jia and W.~L.~Sang,
``Next-to-next-to-leading-order QCD corrections to $e^+e^-\to J/\psi+\eta_c$ at $B$ factories,''
[arXiv:1901.08447 [hep-ph]].
%15 citations counted in INSPIRE as of 02 Jun 2022


\bibitem{gxz1}
G.~Chen, X.~G.~Wu, Z.~Sun, S.~Q.~Wang and J.~M.~Shen,
``Exclusive charmonium production from $e^+ e^-$ annihilation round the $Z^0$ peak,''
Phys. Rev. D \textbf{88}, 074021 (2013),
%doi:10.1103/PhysRevD.88.074021
[arXiv:1308.5375 [hep-ph]].
%6 citations counted in INSPIRE as of 02 Jun 2022

%\cite{Likhoded:2017jmx}
\bibitem{Likhoded:2017jmx}
A.~K.~Likhoded and A.~V.~Luchinsky,
``Double Charmonia Production in Exclusive $Z$ Boson Decays,''
Mod. Phys. Lett. A \textbf{33}, no.14, 1850078 (2018),
%doi:10.1142/S0217732318500785
[arXiv:1712.03108 [hep-ph]].
%12 citations counted in INSPIRE as of 07 Aug 2022

%\cite{Berezhnoy:2021tqb}
\bibitem{Berezhnoy:2021tqb}
A.~V.~Berezhnoy, I.~N.~Belov, S.~V.~Poslavsky and A.~K.~Likhoded,
``One-loop corrections to the processes e+e-$\to$\ensuremath{\gamma}, Z$\to$J/\ensuremath{\psi}\,\ensuremath{\eta}c and e+e-$\to$Z$\to$J/\ensuremath{\psi}\,J/\ensuremath{\psi},''
Phys. Rev. D \textbf{104}, no.3, 034029 (2021),
%doi:10.1103/PhysRevD.104.034029
[arXiv:2101.01477 [hep-ph]].
%2 citations counted in INSPIRE as of 07 Aug 2022

%\cite{Luo:2022ugd}
\bibitem{Luo:2022ugd}
X.~Luo, H.~B.~Fu, H.~J.~Tian and C.~Li,
``Next-to-leading-order QCD correction to the exclusive double charmonium production via $Z$ decays,''
[arXiv:2209.08802 [hep-ph]].
%0 citations counted in INSPIRE as of 07 Oct 2022

%\cite{Berezhnoy:2016etd}
\bibitem{Berezhnoy:2016etd}
A.~V.~Berezhnoy, A.~K.~Likhoded, A.~I.~Onishchenko and S.~V.~Poslavsky,
``Next-to-leading order QCD corrections to paired $B_c$ production in $e^+e^-$ annihilation,''
Nucl. Phys. B \textbf{915}, 224-242 (2017),
%doi:10.1016/j.nuclphysb.2016.12.013
[arXiv:1610.00354 [hep-ph]].
%12 citations counted in INSPIRE as of 07 Aug 2022

%\cite{Belov:2021ftc}
\bibitem{Belov:2021ftc}
I.~N.~Belov, A.~Berezhnoy and E.~Leshchenko,
``Associated Charmonium-Bottomonium Production in a Single Boson e+e\ensuremath{-} Annihilation,''
Symmetry \textbf{13}, no.7, 1262 (2021),
%doi:10.3390/sym13071262
[arXiv:2105.06174 [hep-ph]].
%0 citations counted in INSPIRE as of 07 Aug 2022

%%%%%%%%%%%%%%%%%%%%%%%%%%%%
%FORMULATION
%%%%%%%%%%%%%%%%%%%%%%%%%%%%

\iffalse
%octet matrix-suppressed
%\cite{Wu:2002ig}
\bibitem{Wu:2002ig}
X.~G.~Wu, C.~H.~Chang, Y.~Q.~Chen and Z.~Y.~Fang,
``The Meson $B_c$ annihilation to leptons and inclusive light hadrons,''
Phys. Rev. D \textbf{67}, 094001 (2003),
%doi:10.1103/PhysRevD.67.094001
[arXiv:hep-ph/0209125 [hep-ph]].
%22 citations counted in INSPIRE as of 22 Sep 2021

%octet matrix-fit
%\cite{Ma:2010yw,Ma:2010jj}
\bibitem{Ma:2010yw}
Y.~Q.~Ma, K.~Wang and K.~T.~Chao,
``$J/\psi (\psi^\prime)$ production at the Tevatron and LHC at ${\cal O}(\alpha_s^4v^4)$ in nonrelativistic QCD,''
Phys. Rev. Lett. \textbf{106}, 042002 (2011),
%doi:10.1103/PhysRevLett.106.042002
[arXiv:1009.3655 [hep-ph]].
%\cite{Ma:2010jj}
\bibitem{Ma:2010jj}
Y.~Q.~Ma, K.~Wang and K.~T.~Chao,
``A complete NLO calculation of the $J/\psi$ and $\psi^\prime$ production at hadron colliders,''
Phys. Rev. D \textbf{84}, 114001 (2011),
%doi:10.1103/PhysRevD.84.114001
[arXiv:1012.1030 [hep-ph]].
\fi


%%%%%%%%%%%%%%%%%%%%%%%%%%%%%%
%NUMERICAL ANALYSIS
%%%%%%%%%%%%%%%%%%%%%%%%%%%%%%
% Liao's estimates of wave fucntions
\bibitem{lx}
Q.~L.~Liao and G.~Y.~Xie,
``Heavy quarkonium wave functions at the origin and excited heavy quarkonium production via top quark decays at the LHC,''
Phys. Rev. D \textbf{90}, no.5, 054007 (2014),
%doi:10.1103/PhysRevD.90.054007
[arXiv:1408.5563 [hep-ph]].
%7 citations counted in INSPIRE as of 02 Jun 2022


%PDG
\bibitem{pdg}
M.~Tanabashi \textit{et al.} [Particle Data Group],
``Review of Particle Physics,''
Phys. Rev. D \textbf{98}, no.3, 030001 (2018)
%doi:10.1103/PhysRevD.98.030001
%7800 citations counted in INSPIRE as of 02 Jun 2022

%\cite{CMS:2022fsq}
\bibitem{CMS:2022fsq}
 [CMS],
``Search for Higgs boson decays into Z and J/$\psi$ and for Higgs and Z boson decays into J/$\psi$ or $\Upsilon$ pairs in pp collisions at $\sqrt{s}$ = 13 TeV,''
[arXiv:2206.03525 [hep-ex]].
%0 citations counted in INSPIRE as of 12 Jun 2022

%\cite{FCC:2018evy}
\bibitem{FCC:2018evy}
A.~Abada \textit{et al.} [FCC],
``FCC-ee: The Lepton Collider: Future Circular Collider Conceptual Design Report Volume 2,''
Eur. Phys. J. ST \textbf{228}, no.2, 261-623 (2019).
%doi:10.1140/epjst/e2019-900045-4
%662 citations counted in INSPIRE as of 09 Oct 2022




%%%%%%%%%%%%%%%%%%%%%%%%%%%%%%
%SUMMARY
%%%%%%%%%%%%%%%%%%%%%%%%%%%%%%

\end{thebibliography}
\end{document}